\documentclass[preprint,pre,aps]{revtex4}
\usepackage[dvips]{graphicx}
\usepackage{color}
\bibstyle{prsty}
\begin{document}
\title{%Phase diagram and EOS in systems with competing interactions from a mesoscopic approach 
Mesoscopic theory for systems with competing interactions near a confining wall}
%Density functional theory for self-assembling systems with long-wavelength fluctuations taken into account}
\author{ A. Ciach}
\affiliation{Institute of Physical Chemistry,
 Polish Academy of Sciences, 01-224 Warszawa, Poland}
  \date{\today} 
\begin{abstract}
Mesoscopic theory for self-assembling systems near a planar confining surface is developed. Euler-Lagrange (EL) equations 
and the boundary conditions (BC) for the local volume fraction and the correlation function are derived from the
DFT expression for the grand thermodynamic potential. Various levels of approximation can be considered for the obtained equations.
The lowest-order nontrivial approximation (GM) resembles the Landau-Brazovskii type theory for a semiinfinite system. 
Unlike in the original phenomenological theory, however, all coefficients 
 in our equations and BC are expressed in terms of the interaction potential and the thermodynamic state. Analytical solutions
 of the linearized equations in GM are presented and discussed on a general level and for a particular example
 of the double-Yukawa potential. We show exponentially damped oscillations of the volume fraction and the correlation function
 in the direction perpendicular to the confining surface. The correlations show oscillatory decay in directions
 parallel to this surface too, with the decay length increasing  significantly when the system boundary is  approached.
  The framework of our theory allows for 
 a systematic improvement of the accuracy of the results. 
\end{abstract}
\maketitle
\section{Introduction}
Competing interactions may lead to a self-assembly into different aggregates that at sufficiently low temperature $T$
form periodic 
patterns on a mesoscopic length scale~\cite{seul:95:0,matsen:96:1,stradner:04:0,imperio:04:0,candia:06:0,archer:07:1,
ciach:08:1,ciach:10:1,ciach:13:0,lindquist:16:0,pini:17:0,edelmann:16:0,zhuang:16:0,zhuang:16:1,sweatman:14:0}.
The sequence of the ordered phases at low $T$ is the same in block copolymer systems~\cite{matsen:96:0} 
and in systems containing particles interacting with the effective  SALR potential that is  attractive at short-
and  repulsive at large separations 
~\cite{ciach:13:0,pini:17:0,edelmann:16:0,zhuang:16:0}. 
This universal behavior follows from the fact that  on a qualitative level 
both systems can be described by the Landau-Brazovskii functional (LB)~\cite{ciach:08:1,ciach:13:0}. A typical  
example of the SALR potential is the effective interaction between charged spherical particles 
(colloidal particles, nanoparticles or globular proteins)
in a solvent inducing strong short-range
effective attraction between them~\cite{stradner:04:0,zhang:09:0,campbell:05:0,royall:18:0,bergman:19:0}. One should mention,
however that in the SALR systems the ordered phases have not been observed experimentally yet~\cite{royall:18:0}.
On the other hand, the cubic phases that are only metastable
in the LB theory, turned out to be stable in multicomponent mixtures containing surfactant or 
lipids~\cite{pieranski:00:0,pieranski:01:0,pieranski:11:0,latypova:13:0,gozdz:16:0}.
Thanks to this universality, one can be guided by the properties of one system in studies of 
the properties of another system with inhomogeneities on a well-defined length scale, 
even though some deviations from the universal properties of self-assembling systems may exist.
Here we focus on the SALR model that is particularly simple, because the solvent is treated as a structureless medium, 
and only one kind of particles with isotropic interactions need to be considered. 

Because of the periodic structure on the mesoscopic length scale (a few or a few tens of particle diameter),
confinement by solid surfaces or by interfaces may lead to significant structural changes. The structural transformations 
depend on the properties of the system boundaries, and on the compatibility between
the symmetry and period of the ordered structure and the shape and size of the confinement.
The effect of confinement on the structure,
mechanical and thermal properties of the ordered phases in amphiphilic and SALR systems was studied by 
theory, simulations and experiment~\cite{matsen:97:0,shi:13:0,doerk:17:0,tasinkevych:01:0,tasinkevych:05:0,ciach:11:0,archer:08:0,latypova:14:0,hu:18:0,imperio:07:0,almarza:16:0,serna:19:0,pekalski:19:0,antelmi:95:0,kekicheff:97:0,pieranski:11:0,latypova:13:0,richardson:14:0}. 
In particular, structures absent in the bulk can be induced by appropriate
boundaries~\cite{gozdz:15:0,pekalski:19:0,serna:19:0}. 

Less attention was paid to effects of a single planar wall on the disordered phase in the soft-matter 
systems~\cite{ciach:99:0,archer:16:0,litniewski:19:0,bildanau:19:0}. 
In the disordered phase the density is position-independent, 
and the correlation function  exhibits either a monotonic or an oscillatory decay, just as in the case of simple fluids, but
the wavelength of the oscillatory 
decay is set by the ranges of the attractive and repulsive parts of the interactions in the SALR systems, 
not by the size of the particles. The snapshots, however, show that the structure
of the disordered phase in simple fluids and in the SALR systems can be completely different.
At low volume fractions $\zeta$ and/or high $T$, the particles are more or 
less homogeneously distributed in both cases.
When a so called critical cluster concentration line in the $(\zeta,T)$ phase diagram is crossed in the SALR system,
however, clusters with a well-defined size appear~\cite{santos:17:0,hu:18:0}.
The isolated particles ('monomers') still dominate until another
structural crossover at a higher $\zeta$ is reached. At this crossover, the probability of finding a monomer is equal to the 
probability of finding a particle belonging to a cluster of the optimal size, and
 the specific heat takes a maximum~\cite{litniewski:19:0}. 
 %For larger $\zeta$, the clusters dominate. 
Further increase of the volume fraction leads to another
structural crossover to a percolating network of particles~\cite{zhuang:16:1}. The nontrivial structure of the disordered phase,
in particular the strong inhomogeneities on a well-defined length scale, suggest nontrivial effects of an 
attractive or repulsive surface on the disordered phase in the SALR system.
Indeed, simulations show: (i) formation of inhomogeneous layer of particles  adsorbed at the surface, followed by 
strong depletion of particles in the subsequent layer 
(ii) anomalous decrease of adsorption for increasing chemical potential (iii)
 much larger correlation length  near the confining wall than in the bulk. All these anomalies occur when
 clusters dominate over the monomers~\cite{litniewski:19:0,bildanau:19:0}. 

 Due to the broken 
translational and rotational symmetries, the average volume fraction depends on the distance from the wall, $z$,
and the correlation function between the points ${\bf r}_1$ and  ${\bf r}_2$ depends on $z_1$, $z_2$ and the distance
$r_{\parallel}$ between the projections of the two points on the surface $z=0$. In addition, formation of a structure
with  periodic order 
in the lateral direction in the near-surface region cannot be ruled out~\cite{archer:16:0}. Thus, the problem is very complex.
Because of this complexity, there is a need for an approximate theory
that could give at least qualitative predictions with a reasonable effort. The LB  theory developed for the unconfined SALR 
system in Ref.~\cite{ciach:08:1,ciach:13:0,ciach:10:1,ciach:18:0} correctly predicts the sequence of phases. 
The results can be obtained much more easily than in the standard DFT or liquid theories. 
 Importantly, in the original phenomenological LB theory there is a number of free phenomenological parameters, but
in the theory developed in Ref.~\cite{ciach:08:1,ciach:13:0}, all parameters are expressed in terms of the interaction 
potential, $\zeta$ and $T$. In addition, the high-$T$ part of the phase diagram obtained in 
MF theories is qualitatively different from the simulation results, whereas in this theory it agrees
with simulations on a semi-quantitative level when the fluctuation contribution is added to the grand potential~\cite{ciach:18:0}.

In this work we generalize the LB theory  developed in Ref.~\cite{ciach:08:1,ciach:13:0} to the case of a semiinfinite system.
Broken translational and rotational invariance, however, makes the derivation of the theory more difficult.
The theory is developed in sec.\ref{sec:development}. We limit ourselves to the MF approximation.
In MF, the internal energy of a disordered phase in bulk is approximated by
$U=\frac{1}{2}\rho^2\int V(r) d{\bf r}$, where $\rho$ is the average density and $V(r)$ is the interaction potential. 
This kind of MF approximation is typically made in standard DFT theories. The above expression for $U$ gives the same
internal energy for homogeneous and inhomogeneous systems with the same average density.  However, when clusters of a 
size determined by the range of attraction, separated by a distance determined by the range of repulsion are formed,
the energy is much lower. This is because much more pairs of particles corresponding to a minimum of $V(r)$, 
and much less pairs of particles corresponding to the maximum of $V(r)$ are present than in the case of homogeneously 
distributed particles. As a result, the internal energy obtained by averaging the energy of microstates 
with the proper probability
distribution differs from the energy calculated for the average density. This lower internal energy is associated with
 mesoscopic fluctuations of density or volume fraction in the disordered phase, where the clusters are not localized
 and move almost freely. Let us imagine a fixed mesoscopic part of a system 
 containing clusters or other aggregates. The mesoscopic size means in this context the size comparable with the 
 length scale of inhomogeneities (size of the clusters).
 The local volume fraction (or the number of particles in the mesoscopic window) is significantly larger or
 smaller than the average 
 volume fraction when a cluster enters or leaves the window, respectively. In the case of a homogeneous system, 
 the mesoscopic fluctuations are much smaller.
 What distinguishes the homogeneous and inhomogeneous systems with the same volume fraction, is the variance of 
 the local volume fraction, associated with larger density inside the clusters than between them. 
 The MF theories, including the standard DFT, correctly predict the low- and the high-temperature 
 properties, where either the periodic phases are formed, or the clusters are not yet well developed, 
 and fluctuations of the local volume fraction described above 
 do not play a primary role. The fluctuation contribution to the grand potential will be considered in a forthcoming article
 by combining the DFT and the statistical field theories, as in Ref.~\cite{ciach:18:0}.  

We start in sec.\ref{sec:derivation}  from 
the standard DFT expression for the grand potential
and transform it to an equivalent form that is more suitable for making approximations.
In secs.\ref{sec:EL} and \ref{sec:BC} we derive Euler-Lagrange (EL) equations and the boundary conditions (BC)
from our expression for the grand potential. In sec.\ref{sec:TLst} we 
derive equations for the average volume fraction,
and for the periodic modulations of $\zeta$ in the planes parallel to the wall.
The latter is not considered in the following sections. We focus on the short-range order
as described by the correlation function. In sec.\ref{sec:linEL} we present and discuss linearized EL equations. 
Finally, in sec.\ref{sec:cf} the EL for the correlation function is developed, and various approximate versions are discussed. 
The solutions of the obtained EL equations are in principle equivalent to the results of minimization of the functional 
that was a starting point of our derivation. However, these equations can be greatly simplified by following the steps
leading to the LB theory in the bulk~\cite{ciach:08:1,ciach:10:1,ciach:13:0}. 
In sec.\ref{sec:gm} such a generic model is developed for a semiinfinite system.
The EL equation for the volume 
fraction (sec.\ref{sec:vfgm}) and the correlation function (sec.\ref{sec:cfgm}) 
as well as the BC take a particularly simple form. The linearized equations
can be easily solved analytically, and we discuss properties of these solutions on a general level.
In sec.\ref{sec:example} our theory is applied to a double-Yukawa  interaction potential. 
The shapes of the excess volume fraction and
the correlation function are presented and discussed. We summarize in sec.\ref{section:sum}.

\section{\label{sec:development}Development of the theory}
In this work we consider the effect of the wall on the local structure when the disordered phase is stable in the bulk, and 
far away from the confining surface, the volume fraction $\zeta_b$ of the particles is position-independent.
We assume that the confining $(x,y)$ plane is at $z=0$, and the particles $z$-th coordinate is $z\ge 0$. 
We develop a mean-field (MF) theory and assume that the grand potential can be written in the form
\begin{equation}
\label{Omega}
 \beta\Omega=\beta U +\beta U_{ext} +\int d{\bf r}_{\parallel} \int_0^{\infty} dz\beta f_h(\zeta({\bf r}_{\parallel},z))
 -\beta\mu \int d{\bf r}_{\parallel} \int_0^{\infty} dz \zeta({\bf r}_{\parallel},z) 
\end{equation}
where $\mu$  is the chemical potential, $\beta=1/(k_BT)$, and $k_B$ is the Boltzmann factor. 
The third term in Eq.(\ref{Omega}) is the entropic contribution. 
In a popular approach, $f_h$ is the hard-core reference-system free energy in the local-density approximation.
The first two terms in (\ref{Omega})  are the contributions to the internal energy associated with
interparticle interactions and an external potential, 
respectively, and are given by
\begin{equation}
\label{U}
 U[\zeta]=\int d{\bf r}_{\parallel}\int_0^{\infty} dz\int d\Delta{\bf r}_{\parallel}\int_0^{\infty} 
 d\Delta z \zeta({\bf r}_{\parallel},z)V(\Delta r_{\parallel},\Delta z) 
 \zeta({\bf r}_{\parallel}+\Delta{\bf r}_{\parallel},z+\Delta z)
\end{equation}
and 
\begin{equation}
\label{Usurf}
 U_{ext}=\int d{\bf r}_{\parallel}\int_0^{\infty} dz V_{ext}({\bf r}_{\parallel}, z) \zeta({\bf r}_{\parallel},z).
\end{equation}
 In (\ref{U}) and (\ref{Usurf}), $V$ and $V_{ext}$ denote the interparticle interactions,
and the interactions between the particles and the wall, respectively. In the case of the 
structureless wall to which we will restrict our attention later, $V_{ext}$ depends only on $z$.
We assume that the interparticle interactions are spherically symmetric, and depend only on $r=\sqrt{r_{\parallel}^2+z^2}$,
but for convenience we will consider the directions parallel and perpendicular to the confining surface separately.

We introduce the excess grand potential 
\begin{equation}
 \Delta \beta\Omega[\zeta]=\beta\Omega[\zeta]-\beta\Omega_b[\zeta_b],
\end{equation}
where $\Omega_b[\zeta_b]$ is the grand potential of the considered system with the same volume in the bulk.
The volume fraction in the bulk, $\zeta_b$, satisfies the equation
\begin{equation}
\label{bulkzeta}
 \zeta_b \int d\Delta{\bf r}_{\parallel}\int_{-\infty}^{\infty} d\Delta z \beta V(\Delta r_{\parallel},\Delta z) 
 + \beta f_h^{'} =\beta\mu,
\end{equation}
that follows from the minimization of the grand potential of the unconfined system with
the position-independent volume fraction. Here and below,  $f^{'} (\zeta)=df/d\zeta$ is used for any function $f$.
In the next subsection we transform $\Delta \beta\Omega[\zeta]$ to a form more convenient for  approximations  and
analytical solutions.

\subsection{\label{sec:derivation}Derivation of a new version of the density functional}

The local volume fraction can be split into the bulk and the excess terms,
\begin{equation}
 \zeta({\bf r}_{\parallel},z)=\zeta_b + \Delta\zeta({\bf r}_{\parallel},z), 
\end{equation}
where ${\bf r}_{\parallel}$ is a two-dimensional vector 
in a plane parallel to the confining wall.  In general, we do not exclude the possibility of the wall-induced 
long-range order in the directions parallel to the wall. By the long-range order in
the directions parallel to the wall
we mean a periodic structure in the $(x,y)$ plane. If the surface induces only short-range ordering 
reflected in the oscillatory decay of the correlation function in the  $(x,y)$ plane,
but not on the level of the one-point volume fraction or density,
then $\Delta\zeta({\bf r}_{\parallel},z)$
is independent of ${\bf r}_{\parallel}$.
In the directions parallel to the surface, either translational invariance or periodic structure can be expected,
and it is convenient to consider the volume fraction and the interaction potential 
in the mixed Fourier- and real-space representation,
\begin{equation}
 \tilde \zeta({\bf k}_{\parallel},z)=(2\pi)^2 \delta({\bf k}_{\parallel})\zeta_b
 +\Delta \tilde \zeta({\bf k}_{\parallel},z),
\end{equation}
\begin{equation}
 \Delta \tilde \zeta({\bf k}_{\parallel},z)=\int d{\bf r}_{\parallel} 
 e^{i{\bf k}_{\parallel}\cdot {\bf r}_{\parallel} } \Delta\zeta({\bf r}_{\parallel},z)
\end{equation}
and
\begin{equation}
\label{FTV}
 \tilde V({\bf k}_{\parallel}, \Delta z)=\int d{\bf r}_{\parallel}
 e^{i{\bf k}_{\parallel}\cdot {\bf r}_{\parallel} } V({\bf r}_{\parallel},\Delta z).
\end{equation}
Here and below, tilde denotes a two-dimensional  Fourier transform in the plane parallel to the confining surface, and
$\delta$ denotes the Dirac delta function.

In the mixed representation, the excess grand potential is given by
 \begin{eqnarray}
 \label{DeltaOmga}
   \Delta \beta\Omega[\zeta]&=& \int\frac{d{\bf k}_{\parallel}}{(2\pi)^2}\int_0^{\infty} dz\int_{-\infty}^{\infty} d\Delta z
    \Delta \tilde \zeta(-{\bf k}_{\parallel},z)\beta\tilde V({\bf k}_{\parallel}, \Delta z)\theta(\Delta z)
 \Delta \tilde \zeta({\bf k}_{\parallel},z+\Delta z)
 \\
 \nonumber
&+& \int d{\bf r}_{\parallel}\int_0^{\infty} dz g(\zeta_b, \Delta \zeta) +\beta U_{ext}
 \end{eqnarray}
where $\theta(\Delta z)$  is the Heaviside unit step function,
\begin{eqnarray}
\label{g}
 g(\zeta_b, \Delta \zeta)=\beta f_h(\zeta_b+\Delta\zeta)-\beta f_h(\zeta_b)
 -\beta f_h^{'}(\zeta_b)\Delta\zeta,
\end{eqnarray}

\begin{equation}
\label{Usurf1}
 U_{ext} =\int d{\bf r}_{\parallel}\int_0^{\infty} dzV_{ext}({\bf r}_{\parallel},z)\zeta({\bf r}_{\parallel},z)
 =\int \frac{d{\bf k}_{\parallel}}{(2\pi)^2}\int_0^{\infty} dz 
 \tilde V_{ext}(-{\bf k}_{\parallel}, z) \tilde \zeta({\bf k}_{\parallel},z),
\end{equation}
and Eq.(\ref{bulkzeta}) was used. 
In the case of a homogeneous confining surface, with no lateral pattern, the 
wall-particle interactions are
independent of $r_{\parallel}$, $V_{ext}({\bf r}_{\parallel},z)=V_{ext}(z)$, and we get from (\ref{Usurf1})
$ U_{ext} =\int_0^{\infty} dz V_{ext}(z)\tilde \zeta({\bf 0},z)$. If, in addition the external potential is
 of very short range,
$ U_{ext}$ can be approximated by 
\begin{equation}
 \label{Usurf2}
  U_{surf} = h \tilde \zeta({\bf 0},0).
\end{equation}

In the rest of this section  we will transform the functional (\ref{DeltaOmga}) to a form that is suitable for 
making approximations based on physical properties of the system. 
Our procedure is similar to the one developed in Ref.~\cite{ciach:13:0,ciach:08:1}
 for the bulk inhomogeneous system.
 
In the first step we consider the internal energy contribution in Eq.(\ref{DeltaOmga}).
The Fourier transform of $ \tilde V_s({\bf k}_{\parallel}, \Delta z)=\tilde V({\bf k}_{\parallel}, \Delta z)\theta(\Delta z)$
in the perpendicular direction contains both the real and the imaginary part 
\begin{equation}
\label{hatV}
\hat V_s({\bf k}_{\parallel},k_{\perp})= 
\int_{-\infty}^{\infty} d\Delta z\tilde V({\bf k}_{\parallel}, \Delta z)  \theta(\Delta z)
 e^{ik_{\perp}\Delta z}=
 \hat V_R({\bf k}_{\parallel},k_{\perp})+i \hat V_I({\bf k}_{\parallel},k_{\perp}).
\end{equation}
In (\ref{hatV}) and below, a three-dimensional Fourier transform is indicated by a hat, to distinguish it form
the two-dimensional Fourier transform indicated by a tilde. 

Let us first focus on the real part of $\hat V_s({\bf k}_{\parallel},k_{\perp})$,
\begin{eqnarray}
 \hat V_R({\bf k}_{\parallel},k_{\perp})=\int_0^{\infty} 
 d\Delta z\tilde V({\bf k}_{\parallel}, \Delta z) \cos(k_{\perp}\Delta z)=
 \frac{1}{2}\int_{-\infty}^{\infty} 
 d\Delta z\tilde V({\bf k}_{\parallel}, \Delta z) e^{ik_{\perp}\Delta z},
\end{eqnarray}
where we used the property $\tilde V({\bf k}_{\parallel}, \Delta z) =\tilde V({\bf k}_{\parallel}, -\Delta z) $.
From the above
and Eq.(\ref{FTV}), we obtain
\begin{eqnarray}
 \hat V_R({\bf k}_{\parallel},k_{\perp})=\frac{1}{2}\int d{\bf r}                                             
 V({\bf r})e^{i{\bf r}\cdot{\bf k}}=\frac{1}{2}\hat U(k^2)
\end{eqnarray}
where ${\bf r}=({\bf r}_{\parallel},z)$,
${\bf k}=({\bf k}_{\parallel},k_{\perp})$,
and we introduced the function $\hat U$ of $k^2=k_{\parallel}^2+k_{\perp}^2$, based on the fact
that the Fourier transform of 
 the interaction potential  is an even function of $k$. 
 We Taylor expand $\hat U(k_{\parallel}^2+k_{\perp}^2)$ in terms of $k_{\perp}^2$,
 \begin{equation}
 \label{VR}
   \hat V_R({\bf k}_{\parallel},k_{\perp})=\frac{1}{2}\sum_{n=0}^{\infty} U_{2n}(k_{\parallel}) k_{\perp}^{2n}.
 \end{equation}
 
In the next step we Fourier-transform $ \hat V_R({\bf k}_{\parallel},k_{\perp})$ given in Eq.~(\ref{VR}) back to the real space
in the direction perpendicular to the surface $z=0$, introduce
the operator $\beta\hat U\Big(k_{\parallel}^2-\frac{\partial^2}{\partial z^2}\Big)$
acting on $\Delta\tilde \zeta$ according to the equation
\begin{eqnarray}
\label{Uaction}
 \hat U\Big(k_{\parallel}^2-\frac{\partial^2}{\partial z^2}\Big)\Delta\tilde \zeta ({\bf k}_{\parallel}, z)
&=& \sum_{n=0}^{\infty}(-1)^n U_{2n}(k_{\parallel} ) 
 \frac{\partial^{2n}\Delta \tilde \zeta ({\bf k}_{\parallel},z)}{\partial z^{2n}},
\end{eqnarray}
 and  obtain the corresponding contribution to the internal energy 
\begin{eqnarray}
 \beta \Delta U_R=\frac{1}{2}\int\frac{d{\bf k}_{\parallel}}{(2\pi)^2}  
 \int_0^{\infty}dz \Delta\tilde \zeta (-{\bf k}_{\parallel}, z)
 \beta\hat U\Big(k_{\parallel}^2-\frac{\partial^2}{\partial z^2}\Big)\Delta\tilde \zeta ({\bf k}_{\parallel}, z).
\end{eqnarray}

Let us now focus  on the imaginary part  $V_I$
that is an odd function of $k_{\perp}$,
\begin{eqnarray}
\label{VI}
 \hat V_I({\bf k}_{\parallel},k_{\perp})=\int_0^{\infty} 
 d\Delta z\tilde V({\bf k}_{\parallel}, \Delta z) \sin(k_{\perp}\Delta z)=
 \sum_{n=0}^{\infty}I_{2n+1}(k_{\parallel})k_{\perp}^{2n+1},
\end{eqnarray}
where 
\begin{eqnarray}
\label{I2n+1}
 I_{2n+1}(k_{\parallel})=\int_0^{\infty} d\Delta z \tilde V(k_{\parallel},\Delta z)
 \frac{(-1)^n\Delta z^{2n+1}}{(2n+1)!}.
\end{eqnarray}
We Fourier-transform $ \hat V_I({\bf k}_{\parallel},k_{\perp})$ given by (\ref{VI}) back to the real space
in the direction perpendicular to the surface $z=0$, introduce
the operator $\hat V_I(k_{\parallel},i\frac{\partial}{\partial z})$
%which action  on $\Delta\tilde \zeta ({\bf k}_{\parallel}, z)$ is given 
by
\begin{eqnarray}
\label{VI1}
 \hat V_I\Big(k_{\parallel},i\frac{\partial}{\partial z}\Big)\Delta\tilde\zeta({\bf k}_{\parallel}, z)
 = \sum_{n=0}^{\infty} (-1)^n I_{2n+1}(k_{\parallel})\frac{\partial^{2n+1} 
 \Delta\tilde\zeta({\bf k}_{\parallel}, z)}{\partial z^{2n+1}},
\end{eqnarray}
and obtain the following expression for the corresponding contribution to the internal energy
\begin{eqnarray}
\label{betaUI}
  \beta U_I=  \int\frac{d{\bf k}_{\parallel}}{(2\pi)^2}
 \int_0^{\infty} dz \Delta\tilde\zeta(-{\bf k}_{\parallel},z)
 \hat V_I\Big(k_{\parallel},i\frac{\partial}{\partial z}\Big)\Delta\tilde\zeta({\bf k}_{\parallel},z).
\end{eqnarray}
As seen from (\ref{VI1}), $\beta  U_I$ is real as it should be. Because of the odd derivatives in Eq.(\ref{VI1}), however,
Eq.(\ref{betaUI}) reduces after integration by parts to a surface
term of the form
\begin{eqnarray}
\label{bDUI}
 \beta  U_I=-\frac{1}{2}\int\frac{d{\bf k}_{\parallel}}{(2\pi)^2}
 \sum_{n=0}^{\infty} (-1)^n\beta I_{2n+1}(k_{\parallel})\sum_{m=0}^{2n}
 (-1)^m\Delta\tilde\zeta^{(m)}(-{\bf k}_{\parallel},0)\Delta\tilde\zeta^{(2n-m)}({\bf k}_{\parallel},0),
\end{eqnarray}
where $\Delta\tilde\zeta^{(m)}({\bf k}_{\parallel},0)$ denotes the $m$-th derivative of $\Delta\tilde\zeta$
with respect to its second argument $z$ at $z=0$.

The above mathematical transformations lead to the expression for the excess grand potential 
\begin{eqnarray}
\label{DbO}
 \Delta\beta\Omega[\zeta]&=&\frac{1}{2} \int\frac{d{\bf k}_{\parallel}}{(2\pi)^2} \int_0^{\infty} dz 
 \Delta\tilde \zeta(-{\bf k}_{\parallel},z)\beta \hat U\Big(k_{\parallel}^2-\frac{\partial^2}{\partial z^2}\Big)
  \Delta\tilde \zeta({\bf k}_{\parallel},z)
  \\
 \nonumber
  &+& \int d{\bf r}_{\parallel}\int_0^{\infty} dz g(\zeta_b,\Delta\zeta({\bf r}_{\parallel},z))
  +\beta U_{ext}+ \beta U_I,
\end{eqnarray}
where  $\hat U$ is defined in (\ref{Uaction}),
 $ U_{ext}$ is given in (\ref{Usurf1}) or (\ref{Usurf2}), $g$ is defined in (\ref{g}), 
 and $\beta U_I$ is given in (\ref{bDUI}). The above form is convenient for derivation of
 the Euler-Lagrange (EL) equations.
 
\subsection{\label{sec:EL}Euler-Lagrange equations}

 Minimization of $\Delta\beta\Omega[\zeta]$ given by (\ref{DbO}) leads to the equilibrium volume fraction in our MF theory.
 We follow the standard procedure and require that the part 
 of  $\Delta\beta\Omega[\zeta+\delta\zeta]-\Delta\beta\Omega[\zeta]$ linear in $\delta \tilde \zeta({\bf k}_{\parallel},z)$ 
 vanishes.  The EL equation obtained in this way has the form 
 \begin{eqnarray}
 \label{EL}
\beta \hat U\Big(k_{\parallel}^2-\frac{\partial^2}{\partial z^2}\Big)\Delta\tilde\zeta({\bf k}_{\parallel},z)+
\int d{\bf r}_{\parallel} e^{i{\bf k}_{\parallel}\cdot{\bf r}_{\parallel}} 
g^{(1)}(\zeta_b,\Delta\zeta({\bf r}_{\parallel},z))
+\beta\tilde V_{ext}( {\bf k}_{\parallel},z)=0,
 \end{eqnarray}
where we have introduced the notation
$g^{(n)}(\zeta_b,\Delta\zeta)=\partial^n g(\zeta_b,\Delta\zeta)/\partial\Delta\zeta^n$,
with the derivative taken at the indicated value of the second argument. We  expand  
$g^{(1)}(\zeta_b,\Delta\zeta)$ about $\Delta\zeta=0$, take into account that $g^{(1)}(\zeta_b,0)=0$ (see (\ref{g})), and
after truncating the expansion at the second-order term, obtain 
the linearized EL equation
\begin{eqnarray}
 \label{ELlin}
\beta \hat U\Big(k_{\parallel}^2-\frac{\partial^2}{\partial z^2}\Big)\Delta\tilde\zeta({\bf k}_{\parallel},z)+
A_2(\zeta_b)\Delta\tilde\zeta({\bf k}_{\parallel},z)
+\beta\tilde V_{ext}( {\bf k}_{\parallel},z)=0,
 \end{eqnarray}
where we have introduced $A_n(\zeta_b)=g^{(n)}(\zeta_b,0)=d^n \beta f_h(\zeta_b)/d\zeta_b^n$ for $n\ge 2$ 
to simplify the notation.
In order  to be able to calculate $\Delta\tilde\zeta({\bf k}_{\parallel},z)$, 
it remains to determine the boundary conditions (BC).

\subsection{\label{sec:BC}Boundary conditions}

 In derivation of (\ref{EL}), we have performed integration by parts 
 to get  rid  of the derivatives of $\delta \tilde\zeta$ that appear because of the
 presence of the differential operator in (\ref{DbO}). However,
 in this way additional boundary terms  
 that are proportional to $\delta \tilde\zeta ({\bf k}_{\parallel},0)$ and its  derivatives are generated. 
 In the case of the semiinfinite system, the boundary conditions follow from the requirement that the 
 surface contribution to 
 $\Delta\beta\Omega[\zeta+\delta\zeta]-\Delta\beta\Omega[\zeta]$,
  \begin{eqnarray}
  \label{BC}
  \delta\beta \Omega_{surf}= \sum_{n=1}^{\infty}(-1)^{n}  \Bigg[  \beta I_{2n-1}({\bf k}_{\parallel})\sum_{m=0}^{2n-2}(-1)^m 
 \Delta\tilde\zeta^{(2n-m-2)}({\bf k}_{\parallel},0)
  \delta\tilde\zeta^{(m)}({\bf k}_{\parallel},0)
  \\                      
  \nonumber
  +  \frac{1}{2}\beta U_{2n}({\bf k}_{\parallel})\sum_{m=0}^{2n-1}(-1)^m \Delta\tilde\zeta^{(2n-m-1)}({\bf k}_{\parallel},0)
 \delta\tilde\zeta^{(m)}({\bf k}_{\parallel},0) \Bigg],
 \end{eqnarray}
 coming from $\beta U_I$                
 as well as from the integration by parts mentioned  above, vanishes. 
The first two BC (the terms proportional
to $\delta\tilde \zeta({\bf k}_{\parallel},0)$ and  $\delta\tilde\zeta^{'} ({\bf k}_{\parallel},0))$, are
\begin{eqnarray}
 \label{BC1}
 \sum_{n=1}^{\infty} (-1)^{n}\Bigg[ \frac{1}{2}
 \beta U_{2n}(k_{\parallel}) \Delta\tilde\zeta^{(2n-1)}({\bf k}_{\parallel},0)+
 \beta I_{2n-1}(k_{\parallel}) \Delta\tilde\zeta^{(2n-2)}({\bf k}_{\parallel},0) \Bigg]=0
\end{eqnarray}
and
\begin{eqnarray}
 \label{BC2}
 \sum_{n=1}^{\infty} (-1)^{n} \Bigg[ 
 \beta I_{2n+1}(k_{\parallel}) \Delta\tilde\zeta^{(2n-1)}({\bf k}_{\parallel},0)-
 \frac{1}{2}\beta U_{2n}(k_{\parallel}) \Delta\tilde\zeta^{(2n-2)}({\bf k}_{\parallel},0)\Bigg]=0.
\end{eqnarray}
If (\ref{Usurf2}) instead of (\ref{Usurf1}) is assumed for $U_{ext}$, then the last term in (\ref{EL}) or (\ref{ELlin})
should be removed, and 
$(2\pi)^2\delta({\bf k}_{\parallel})\beta h$ should be added to the LHS of (\ref{BC1}). 
Additional BC are $\lim_{z\to\infty} \Delta \tilde \zeta^{(m)}(k_{\parallel},z)=0$.

\subsection{\label{sec:TLst}Transverse and lateral structure}

The local excess volume fraction in the real-space or in the Fourier representation can be written in the form
\begin{eqnarray}
 \label{zphi}
  \Delta\zeta({\bf r}_{\parallel},z)=\Delta\zeta_0(z)+\phi({\bf r}_{\parallel},z)
\end{eqnarray}
or
\begin{eqnarray}
\label{zphiF}
  \Delta\tilde\zeta({\bf k}_{\parallel},z)=(2\pi)^2\delta({\bf k}_{\parallel}) \Delta\zeta_0(z)
  +\tilde \phi({\bf k}_{\parallel},z).
\end{eqnarray}
By $ \Delta\zeta_0(z)$ we denote the excess volume fraction at the distance $z$ from the surface,
averaged over the $(x,y)$ plane. The functions $\phi({\bf r}_{\parallel},z)$ and
$\tilde \phi({\bf k}_{\parallel},z)$ in turn are associated with the 
long-range lateral order, i.e. with periodic 
density oscillations  in the  plane parallel to the confining wall and separated by the distance  $z$ from it. 
With the above definition, we have
$\int d{\bf r}_{\parallel}\phi({\bf r}_{\parallel},z)=0$ and 
 $\tilde \phi({\bf 0},z)=0$. In a similar way we separate the  external potential into the homogeneous and the oscillatory parts
 \begin{eqnarray}
  \tilde V_{ext}({\bf k}_{\parallel},z)=
  (2\pi)^2\delta({\bf k}_{\parallel}) V_{ext}(z)+ \tilde V_{\parallel}({\bf k}_{\parallel},z)
 \end{eqnarray}
with $\tilde V_{\parallel}({\bf 0},z)=0$.
   
Inserting (\ref{zphiF}) in (\ref{EL}) and separating terms proportional to $\delta({\bf k}_{\parallel})$,
gives us for $k_{\parallel}=0$
\begin{eqnarray}
\label{Ulz}
 \beta \hat U\Big(-\frac{\partial^2}{\partial z^2}\Big)\Delta\zeta_0(z)
+\sum_{n=0}^{\infty}\frac{g^{(n+1)}(\zeta_b,\Delta\zeta_0(z))}{n!A_u} 
\int_{A_u}  d{\bf r}_{\parallel} \phi^n({\bf r}_{\parallel},z)+\beta V_{ext}(z)=0,
\end{eqnarray}
and for $ k_{\parallel}\ne 0$
\begin{eqnarray}
\label{Ulzphi}
 \beta \hat U\Big({\bf k}_{\parallel}^2-\frac{\partial^2}{\partial z^2}\Big)\tilde\phi({\bf k}_{\parallel},z)+
\int d{\bf r}_{\parallel} e^{-i{\bf k}_{\parallel}\cdot {\bf r}_{\parallel}}
g^{(1)}(\zeta_b,\Delta\zeta_0(z)+\phi({\bf r}_{\parallel},z))+\beta \tilde V_{\parallel}({\bf k}_{\parallel},z) =0.
\end{eqnarray}
  The integral in (\ref{Ulz}) is over the area of the unit cell of the periodic structure, $A_u$.
As seen from (\ref{Ulz}) and (\ref{Ulzphi}), the near-surface long-range lateral order and the average density in 
the planes parallel to the surface are coupled.

Depending on a thermodynamic state,
the long-range order near
the confining wall, i.e. a structure periodic in the lateral direction, may or may not be present. 
We shall focus on the latter case, where only short-range lateral order
 is present near the surface. In fact for temperature and density considered in Ref.~\cite{litniewski:19:0},
 only short-range lateral order at the surface 
was observed in MD simulations.
 
 For $\phi=0$, Eq.(\ref{Ulz}) takes the simple form
 \begin{eqnarray}
\label{Ulz0}
 \beta \hat U\Big(-\frac{\partial^2}{\partial z^2}\Big)\Delta\zeta_0(z)
+g^{(1)}(\zeta_b,\Delta\zeta_0(z))+\beta V_{ext}(z)=0.
\end{eqnarray}
Eqs(\ref{DbO})-(\ref{Ulz0})  can be a starting 
point for various approximate theories, based on truncated expansions of $\beta \hat U$, $g^{(1)}$ and/or $\beta U_I$.

\subsection{\label{sec:linEL}Linearized EL equation in the absence of long-range lateral order}

 $g^{(1)}(\zeta_b,\Delta\zeta_0(z))$ in Eq.(\ref{Ulz0}) can be Taylor expanded, and for small values of $\Delta\zeta_0(z)$, 
 the expansion can be truncated
at the first-order term. The linearized Eq.(\ref{Ulz0}) then takes the simple form  
 \begin{eqnarray}
\label{Ulzl}
 \beta \hat U\Big(-\frac{\partial^2}{\partial z^2}\Big)\Delta\zeta_0(z)
+A_2(\zeta_b)\Delta\zeta_0(z)+\beta V_{ext}(z)=0.
\end{eqnarray}
If $V_{ext}(z)=h\delta (z)$, the equation (\ref{Ulzl}) becomes even simpler, and the solution is a sum of
terms proportional to $\exp(i\alpha z)$, where  $\alpha$ is a solution of the equation
 \begin{eqnarray}
 \label{al}
\beta \hat U(\alpha^2) +A_2(\zeta_b)=0.
\end{eqnarray}
 In the disordered phase this equation
has no solutions for real $\alpha$, because  the disordered phase is stable when $\beta \hat U(k_0^2) +A_2(\zeta_b)>0$, 
where $\hat U$ takes a minimum at $k^2=k_0^2$. As the linearized equation can be valid for small $\Delta \zeta_0(z)$, i.e. for 
large $z$, the asymptotic decay is given by the solution $\alpha=i\alpha_0\pm\alpha_1$ with the imaginary part
$\alpha_0>0$ with the smallest magnitude.

The correlation function in the bulk,
$ G(\Delta{\bf r})=\langle\zeta({\bf r}) \zeta({\bf r}+\Delta{\bf r})\rangle-\zeta_b^2$,
  is inversely proportional to the second functional derivative of 
the grand potential with respect to local deviations of the volume fraction from the average value.
In Fourier representation  $\hat G(k)$ is inversely proportional 
 to the LHS of Eq.(\ref{al}) with $\alpha=k$. Poles of  $\hat G(k)$, i.e. zeros of the LHS of  Eq.(\ref{al}), 
determine the decay of correlations in the real space representation. 
 This result shows that the decay length and the period of damped oscillations are the same in the correlation
 function in the bulk 
and in the density profile near a flat wall. This observation was confirmed by simulations of a particular 
version of the SALR model in Ref.\cite{litniewski:19:0}.

There exist two possible cases: (i) $\alpha_1=0$ and $\Delta\zeta_0(z)$ decays monotonically,
 or (ii) $\alpha_1\ne 0$ and an oscillatory decay of $\Delta\zeta_0(z)$ takes place for large $z$.
 The first case concerns $\hat U(k^2)$ that takes the global minimum for $k=0$, 
 and has the  expansion $\hat  U(0)+U_2k^2+...$ with $ \hat U(0)<0$, $U_2>0$.
 The second case concerns $\hat U(k^2)$ that takes  the global minimum for $k=k_0>0$, 
 and has the expansion about the minimum
 $\hat U(k_0^2)+v(k^2-k_0^2)^2+...$.
 In the first case, when the expansion of  $\hat U(k^2)$ is truncated at the term proportional to $k^2$, 
 our theory reduces to the standard Landau theory. In the second case, when $\hat U(k^2)$ is approximated 
 by  
 \begin{equation} 
 \label{hatU}
 \hat U(k^2)= \hat U(k_0^2)+v(k^2-k_0^2)^2+...,
 \end{equation}
 our theory reduces to the Landau-Brazovskii theory. 

 In the rest of the work we limit ourselves to the Brazovskii-type theory.
 
 \subsection{\label{sec:cf}Correlation function }

Let us focus on the correlation function
in the case of no long-range lateral order ($\phi=0$).  In the mixed representation, $\tilde G(k_{\parallel}|z_1,z_2)$  
is the correlation function betweem the volume-fraction waves with the wavelength $k_{\parallel}$ in the longitudinal direction
 in the planes at the separations $z_1$ and $z_2$ from the wall.  $\tilde G(k_{\parallel}|z_1,z_2)$ 
describes also the response in the plane at $z=z_1$ 
 to an oscillatory  perturbation  with the wavelength $ k_{\parallel}$ in the plane at $z=z_2$. 
It is convenient to calculate this function 
from the relation
\begin{eqnarray}
\label{G}
 \tilde G({\bf k}_{\parallel}|z_1,z_2)=
\frac{\delta\Delta\tilde\zeta({\bf k}_{\parallel},z_1)}{\delta (-\beta \tilde V_{ext}({\bf k}_{\parallel},z_2))}.
\end{eqnarray}
In order to obtain an equation for $\tilde G({\bf k}_{\parallel}|z_1,z_2)$, 
we proceed as in the case of the Landau theory for simple fluids (see for example Ref.\cite{diehl:91:0}), and
 perform  functional differentiation of Eq.(\ref{EL}) 
with respect to $-\beta  \tilde V_{ext}$. As a result we  obtain
the equation for the correlation function in the absence of the long-range lateral order and
near a homogeneous wall
\begin{eqnarray}
\label{Dbm0}
 \Big[\beta \hat U\Big(k_{\parallel}^2-\frac{\partial^2}{\partial z_1^2}\Big)+g^{(2)}(\zeta_b,\Delta\zeta_0(z_1))\Big]
 \tilde G(k_{\parallel}|z_1,z_2)=\delta(z_1-z_2),
\end{eqnarray}
where we took into account that  the correlation function depends only on $ k_{\parallel}=|{\bf k}_{\parallel}|$.

BC for $G$ can be obtained  by functional differentiation of the BC for
$\Delta\tilde\zeta({\bf k}_{\parallel},z)$, 
and from (\ref{BC1}) we obtain
\begin{eqnarray}
 \label{BC1G}
 \sum_{n=1}^{\infty} (-1)^{n}\Bigg[ \frac{1}{2}
 \beta U_{2n}(k_{\parallel}) \tilde G^{(2n-1)}(k_{\parallel}|0,z_2) +
 \beta I_{2n-1}(k_{\parallel})  \tilde G^{(2n-2)}(k_{\parallel}|0,z_2) \Bigg]=0,
\end{eqnarray}
where $\tilde G^{(m)}( k_{\parallel}|0,z_2)$ denotes the $m$-th  derivative with respect to $z_1$ at $z_1=0$. 
The BC for $z_1\to\infty$ is  
$\lim_{z_1\to\infty} \tilde G(k_{\parallel}|z_1,z_2)=0$.

Functional differentiation of the linearized equation for $G$ (Eq.(\ref{ELlin}))
with respect to $-\beta  \tilde V_{ext}$, gives
the equation for the correlation function in the Gaussian approximation,
\begin{eqnarray}
\label{Dbm}
 \Big[\beta \hat U\Big(k_{\parallel}^2-\frac{\partial^2}{\partial z_1^2}\Big)+A_2(\zeta_b)\Big]
 \tilde G(k_{\parallel}|z_1,z_2)=\delta(z_1-z_2),
\end{eqnarray}
where $\hat U$  and $A_n$ are defined in Eq.(\ref{Uaction}) and below Eq.(\ref{ELlin}), respectively.
Alternatively, 
Eq.(\ref{Dbm}) can be obtained from (\ref{DbO}) and the analog of the Ornstein-Zernike equation
\begin{eqnarray}
 \label{CG}
  \int dz'
  \tilde C({\bf k}_{\parallel},z_1,z')  \tilde G(k_{\parallel},z',z_2)=\delta(z_1-z_2)
 \end{eqnarray}
where 
\begin{eqnarray}
\label{C}
 \tilde C({\bf k}_{\parallel},z_1,z') =\frac{\delta^2\Delta\beta\Omega}
  {\delta\Delta\tilde\zeta(-{\bf k}_{\parallel},z_1)\delta\Delta\tilde\zeta({\bf k}_{\parallel},z')}.
\end{eqnarray}

Note that Eq.(\ref{Dbm0}) shows that the effect of the volume-fraction profile on the local structure is significant.
However, this effect
cannot be determined on the level of the Gaussian approximation (\ref{Dbm}).
In this approximation, 
$\tilde G(k_{\parallel}|z_1,z_2)$ is independent of the excess volume fraction profile and hence on the 
wall-particle interactions. Thus, the solution of Eq.(\ref{Dbm}) cannot accurately describe the close vicinity of the wall.
Particularly large inaccuracy of $\tilde G(k_{\parallel}|z_1,z_2)$ obtained from  Eq.(\ref{Dbm})
is expected for large wall-particle
interactions leading to large $\Delta\zeta_0$.

We assume small $\Delta\zeta_0$ in (\ref{Dbm0}), i.e. small wall-fluid potential and/or large $z$,
truncate the expansion of $g^{(2)}(\zeta_b,\Delta\zeta_0(z_1))$, 
and  obtain the equation
\begin{eqnarray}
\label{Dbm2}
 \Big[\beta \hat U\Big(k_{\parallel}^2-\frac{\partial^2}{\partial z_1^2}\Big)+A_2(\zeta_b)+
 A_3(\zeta_b)\Delta\zeta_0(z_1)+O(\Delta\zeta_0(z_1)^2)\Big]
 \tilde G(k_{\parallel}|z_1,z_2)=\delta(z_1-z_2).
\end{eqnarray}
If  $ A_3(\zeta_b)\Delta\zeta_0(z_1)$ is small, 
 the term  $ A_3(\zeta_b)\Delta\zeta_0(z_1) \tilde G(k_{\parallel}|z_1,z_2)$ can be treated as a perturbation.
$A_3(\zeta_b)$ takes large values for small and large $\zeta_b$,
but vanishes for the critical volume fraction $\zeta_b\approx 0.129$. Thus, the approximation (\ref{Dbm}) is more accurate
for  $\zeta_b\approx 0.129$, where
lamellar structure is expected at low $T$, than for very dilute systems, where clusters are formed. 
We will limit ourselves to the approximate equation (\ref{Dbm})
that gives the asymptotic decay of correlations at large distances $z$ from the wall, and
can be easily solved analytically.

  \section{\label{sec:gm}The generic model for inhomogeneous systems}
  \label{gm}
  Density waves with the wavenumber $k_0$ corresponding to the minimum of $\hat V(k)$ lead to  the lowest internal energy. 
  The waves associated with significantly larger energy occur with significantly smaller probability.
  We assume, following Ref.\cite{ciach:08:1,ciach:13:0} that the  density waves with the wavenumbers 
  significantly different from $k_0$ are much less probable. If such density waves can be disregarded,
  then $\hat U(k^2)$ can be approximated by (\ref{hatU}). 
  The operator $\hat U( k_{\parallel}^2-\frac{\partial ^2}{\partial z^2})$ defined in Eq.(\ref{Uaction}), in this 
  lowest-order nontrivial approximation
  takes the form 
  \begin{equation}
  \label{UB}
   \hat U\Bigg( k_{\parallel}^2-\frac{\partial ^2}{\partial z^2}\Bigg)=v\frac{\partial ^4}{\partial z^4}
   -2v(k_{\parallel}^2-k_0^2)\frac{\partial ^2}{\partial z^2} +v(k_{\parallel}^2-k_0^2)^2 +\hat U(k_0^2).
  \end{equation}
In the consistent approximation, we truncate the  expansion 
of $\hat V_I({\bf k}_{\parallel},k_{\perp})$ (see (\ref{VI})) at the lowest-order term, and obtain 
 \begin{equation}
 \label{VIB}
  \hat V_I({\bf k}_{\parallel},k_{\perp})=I_1(k_{\parallel})k_{\perp},
 \end{equation}
where $I_1(k_{\parallel})=\int_0^{\infty}dz z \tilde V(k_{\parallel},z)$ (see (\ref{I2n+1})).
 Eqs.(\ref{UB})-(\ref{VIB}) define the generic model (GM) for a semiinfinite 
 system with inhomogeneities at the length scale $2\pi/k_0$. Note that apart from the external field term,
 there is only one surface term in this approximation (see (\ref{bDUI})),
  \begin{equation}
  \beta U_I=-\frac{\beta}{2}\int d{\bf k}_{\parallel} I_1(k_{\parallel})\Delta\tilde\zeta({\bf k}_{\parallel},0)^2.
   \end{equation}
 Moreover, the bulk interactions are characterized by just three numbers, 
 $\hat U(k_0^2), k_0$ and $v$ (see (\ref{hatU})).
 As in the standard Landau theory, $ \beta U_I$ describes the missing-neighbors contribution at the surface. 
 There is an important difference between the simple fluids and the systems with the SALR interactions, however. 
 Namely, $-I_1(0)>0$ for the 
 attractive interactions, but  in the SALR systems $-I_1(0)<0$ is expected when the repulsion is strong enough.
 In the first case the attraction 
 by the particles in the bulk is not compensated by the particles missing for $z<0$, 
 while in the second case the repulsion is not compensated. 
 These unbalanced interactions lead to effective repulsion by the confining wall in simple fluids, and 
 to effective attraction in the SALR systems. This effective attraction to a hard wall,
 due to the missing neighbors
 was indeed observed in simulations~\cite{serna:19:0}.
 In the following, we present the EL equations and the solutions of these equations for the GM.

  \subsection{\label{sec:vfgm}The volume-fraction profile in the generic model}
 \label{vfp}
 When the external potential is
  localized at the surface, the EL equation for the excess density and the BC, Eqs.(\ref{Ulz0})
  and (\ref{BC1})-(\ref{BC2}), simplify to 
\begin{eqnarray}
\label{Ulz1}
\beta\Big( v\frac{\partial ^4}{\partial z^4}
   +2vk_0^2\frac{\partial ^2}{\partial z^2} +vk_0^4 +\hat U(k_0^2)\Big)
   \Delta\zeta_0(z)
+g^{(1)}(\zeta_b,\Delta\zeta_0(z))=0,
\end{eqnarray}
 \begin{eqnarray}
 \Big[\frac{ v}{2}\frac{\partial^3}{\partial z^3}+
 v k_0^2 \frac{\partial}{\partial z}- I_1(0)\Big]\Delta\zeta_0(z)|_{z=0}+h=0,
 \end{eqnarray}
 and
 \begin{eqnarray}
 \Big[\frac{ v}{2}\frac{\partial^2}{\partial z^2}+ 2k_0^2 \Big]\Delta\zeta_0(z)|_{z=0}=0.
 \end{eqnarray}
 
 The  linearized equation (\ref{Ulz1})  is simply
 \begin{eqnarray}
\label{Ulz2}
\Big( \frac{\partial ^4}{\partial z^4}
   +2k_0^2\frac{\partial ^2}{\partial z^2} +D_0\Big)
   \Delta\zeta_0(z)=0,
\end{eqnarray}
where
\begin{eqnarray}
\label{D0}
  D_0=\tau^2+ k_0^4
 \end{eqnarray}
 with
 \begin{eqnarray}
 \label{kappa}
 \tau^2=(\hat U(k_0^2)+k_BTA_2(\zeta_b) )/ v.
 \end{eqnarray}
 Here we limit ourselves to the linearized EL equation  (\ref{Ulz2}).
The stability  condition of the disordered phase, $\beta\hat U(k_0^2)+A_2(\zeta_b)>0$, implies that $D_0>0$.
Since $D_0>0$ in the disordered phase,
 the solution of (\ref{Ulz2}) in this phase is
  \begin{eqnarray}
  \label{profile}
   \Delta\zeta_0(z)={\cal A} e^{-\alpha_0 z} \cos(\alpha_1 z +\vartheta)
  \end{eqnarray}
where 
\begin{eqnarray}
\label{alpha}
  \alpha_{0,1}=\Bigg[\frac{1}{2}\Big(\sqrt{D_0}
  \mp k_0^2\Big)\Bigg]^{1/2},
\end{eqnarray}
 \begin{eqnarray}
 \label{calA}
 {\cal A} =-\frac{2\beta h\sqrt{D_0}}{\alpha_1(vD_0-4\alpha_0I_1(0))},
\end{eqnarray}
and
 \begin{eqnarray}
  \tan(\vartheta)=-\frac{k_0^2}{\tau}.
 \end{eqnarray}

 The oscillatory decay (\ref{profile})  at sufficiently large separations 
 is quite universal for systems with inhomogeneities at the 
 well-defined length scale (here $2\pi/k_0$). The formula (\ref{profile})
fits quite well the simulation results for a particular version of the SALR interactions
already for $z>2\pi/\alpha_1$ \cite{litniewski:19:0}. 
The same behavior was predicted for 
the asymptotic decay of the charge density near a charged wall in ionic systems~\cite{ciach:18:1}.
In this case, Eq.(\ref{profile}) fits very well simulation results  already for $z>2\pi/\alpha_1$, too ~\cite{otero:18:0}. 
The values of the parameters in (\ref{profile}), however, agree with simulations performed in Ref~\cite{litniewski:19:0}.
only semi-quantitatively. 
A better agreement with simulations was obtained when $\alpha$ was calculated form Eq.(\ref{al}), 
and the effect of clustering on the entropy was taken into account~\cite{litniewski:19:0}. 
Quantitative agreement, however, cannot be expected, because (i) the gradient expansion restricts the results 
to long-ranged features, (ii) the presented theory is of MF type, and (iii)
the entropic contribution is based on the reference system of hard spheres in the local density approximation. 

 \subsection{\label{sec:cfgm}Correlation function in the generic model}
 
 In the GM the EL equation and the BC for $\tilde G$ are (see Eqs.(\ref{Dbm0}) and (\ref{BC1G}))
 \begin{eqnarray}
 \label{ELGM}
 \beta \Bigg[ v\frac{\partial ^4}{\partial z_1^4}
   +2v (k_0^2-k_{\parallel}^2)\frac{\partial ^2}{\partial z_1^2}  + d(z_1,k_{\parallel})
   \Bigg]\tilde G(k_{\parallel}|z_1,z_2)
   =\delta(z_1-z_2)
\end{eqnarray}
 where $d(z_1,k_{\parallel})=\hat U(k_0^2)+v(k_{\parallel}^2-k_0^2)^2   +k_BTg^{(2)}(\zeta_b,\Delta\zeta_0(z_1))$
and
\begin{eqnarray}
 \Bigg[
 \frac{v}{2}\frac{\partial^3}{\partial z^3} -v(k_{\parallel}^2-k_0^2)\frac{\partial}{\partial z}-I_1(k_{\parallel})\Bigg]
 \tilde G(k_{\parallel}|z,z')|_{z=0}=0.
\end{eqnarray}

 In the Gaussian approximation, Eq.(\ref{ELGM})
 simplifies to (see Eq.(\ref{Dbm}))
\begin{eqnarray}
\label{ELGmm}
 \beta v\Bigg[ \frac{\partial ^4}{\partial z_1^4}
   +2 (k_0^2-k_{\parallel}^2)\frac{\partial ^2}{\partial z_1^2} + 
   D(k_{\parallel}^2)\Bigg]\tilde G(k_{\parallel}|z_1,z_2)=\delta(z_1-z_2)
\end{eqnarray}
where
\begin{eqnarray}
\label{D}
  D(k_{\parallel}^2)=\tau^2+ (k_{\parallel}^2-k_0^2)^2,
 \end{eqnarray}
 and $\tau$ is defined in (\ref{kappa}).
Because in the disordered phase $D(k_{\parallel}^2)>0$,
 the solution of (\ref{ELGmm}) should have the form
\begin{eqnarray}
\label{GAnsatz}
 \tilde G(k_{\parallel}|z_1,z_2)=a_-(k_{\parallel}) e^{-\alpha(k_{\parallel}) |z_1-z_2|}
 +a_-^* (k_{\parallel})e^{-\alpha^*(k_{\parallel}) |z_1-z_2|}
 \\
 \nonumber
 +a_+(k_{\parallel}) e^{-\alpha(k_{\parallel}) (z_1+z_2)} +a_+^*(k_{\parallel}) e^{-\alpha^*(k_{\parallel}) (z_1+z_2)} 
\end{eqnarray}
where $\alpha(k_{\parallel})= \alpha_0(k_{\parallel})+i\alpha_1(k_{\parallel})$, with

\begin{eqnarray}
\label{alphakpar}
  \alpha_{0,1}(k_{\parallel})=\Bigg[\frac{1}{2}\Big(\sqrt{D(k_{\parallel}^2)}
  \pm (k_{\parallel}^2-k_0^2)\Big)\Bigg]^{1/2}.
\end{eqnarray}
The function (\ref{GAnsatz}) with (\ref{alphakpar}) satisfies Eq.(\ref{ELGmm}) provided that 
\begin{eqnarray}
 a_-(k_{\parallel})=-\frac{k_BT}{2v(\alpha^2(k_{\parallel})-\alpha^{*2}(k_{\parallel}))\alpha(k_{\parallel})}.
\end{eqnarray}
The BC determines the amplitude $a_+$,
\begin{eqnarray}
 a_+(k_{\parallel})=\frac{
 v\sqrt{D(k_{\parallel})}\alpha^*(k_{\parallel})+2I_1(k_{\parallel})}{
 v\sqrt{D(k_{\parallel})}\alpha^*(k_{\parallel})-2I_1(k{\parallel})}a_-(k_{\parallel}).
\end{eqnarray}
$ \tilde G(k_{\parallel}|z_1,z_2)$ 
exhibits an oscillatory decay as a function of $|z_1-z_2|$ and $z_1+z_2$ with the same characteristic lengths in both cases,
and Eq.(\ref{GAnsatz}) can be written in the equivalent form
\begin{eqnarray}
\label{Gkzz}
 \tilde G(k_{\parallel}|z_1,z_2)=
 {\cal A}_-(k_{\parallel})e^{-\alpha_0(k_{\parallel}) |z_1-z_2|}\cos[\alpha_1(k_{\parallel}) |z_1-z_2|+\theta_-(k_{\parallel})]
 \\
 \nonumber
 + {\cal A}_+(k_{\parallel})e^{-\alpha_0(k_{\parallel}) (z_1+z_2)}\cos\big[\alpha_1(k_{\parallel}) (z_1+z_2)
 +\theta_+(k_{\parallel})\big].
\end{eqnarray}
The involved dependence of the parameters ${\cal A}_-,{\cal A}_+,\theta_-,\theta_+$ on $k_{\parallel}$ 
will not be given  here.

The inverse decay length $\alpha_0( k_{\parallel})$ is an increasing function of $k_{\parallel}$,
and takes the smallest value $\alpha_0$ for $ k_{\parallel}=0$. Thus, 
in the $z$ direction the correlations between the $ k_{\parallel}=0$ modes decay most slowly, and in the same way as 
$\Delta\zeta_0(z)$. In contrast, the wavelength of oscillations in the $z$ direction, $2\pi/\alpha_1( k_{\parallel})$,
increases with increasing $k_{\parallel}$.
For $ k_{\parallel}<k_0$, we have 
$\alpha_0( k_{\parallel})<\alpha_1( k_{\parallel})$, i.e. 
pronounced oscillations of the correlation function in the transverse direction, 
while $\alpha_0( k_{\parallel})>\alpha_1( k_{\parallel})$
for  $ k_{\parallel}>k_0$ (strongly damped oscillations in the transverse direction).

 The first term in (\ref{Gkzz}) depends only on the separation between the two parallel planes,
and is independent of the distance from the wall. The Fourier transform of this term 
in the $z$ direction
 gives the bulk correlation function in Fourier representation. For $z_1=z_2\gg 1$, i.e. 
 in a single plane away from the wall, Eq.(\ref{Gkzz}) takes a maximum for
$ k_{\parallel}^2=k_0^2-\tau/\sqrt 3$.

For finite $z_1$, $z_2$ the second term in  Eq.(\ref{Gkzz}), describing the effect of the confining wall, becomes important. 
The dependence of the amplitude and the phase  on $k_{\parallel}$ is quite complex,
and depends on $I_1(k_{\parallel})$ that
in turn depends on the shape of the interaction potential. 
We shall discuss $\tilde G(k_{\parallel}|z_1,z_2)$ in more detail for a particular form of $V(r)$ in the next section.

\section{\label{sec:example}Generic model results for double-Yukawa potential}

As an example we consider the popular double-Yukawa potential,

\begin{equation}
\label{int_pot_r}
V(r)=-\frac{K_1}{r}e^{-\kappa_1r}+\frac{K_2}{r}e^{-\kappa_2r}.
\end{equation}
%%%
 For the reference-system
 free-energy density we choose the Percus-Yevick approximation 
 \begin{eqnarray}
\label{PY}
\beta f_h(\zeta)=\rho\ln(\rho)-\rho+
\rho\Bigg[\frac{3\zeta(2-\zeta)}{2(1-\zeta)^2}-\ln(1-\zeta)\Bigg],
\end{eqnarray}
where $\rho=6\zeta/\pi$.
 
 In order to calculate the excess volume fraction and the correlation function, we need to express
 $k_0$, $\hat U(k_0)$, $v$ and $I_1(k_{\parallel})$ in terms of  $K_1$, $K_2$, $\kappa_1$ and $\kappa_2$. 
 For the chosen potential, we can easily obtain analytical expressions (see Appendix). 
 
 We choose the parameters $K_1=1, K_2=0.2,\kappa_1=1,\kappa_2=0.5$ as in
 earlier works focused on the bulk properties~\cite{ciach:10:1,ciach:13:0}. 
 $K_1$ sets the energy unit, and we introduce dimensionless temperature $T^*=k_BT/K_1$.
 For this potential, large clusters are formed, since $\pi/k_0\approx 5$.
 In this lowest-order approximation, the dependence on the thermodynamic state is only through the single parameter $\tau$,
which in turn depends on $k_BTA_2(\zeta_b)$ (see (\ref{kappa})).
In MF, the instability with respect to periodic ordering occurs below the $\lambda$-line given by $\tau=0$. 
Beyond MF, however, such an instability is not present, therefore we calculate 
  the excess volume-fraction profile and the correlation function mainly for 
  $\tau=0.23$ that is well above the $\lambda$-line.
  
\begin{figure}
 \centering
\includegraphics[scale =0.45]{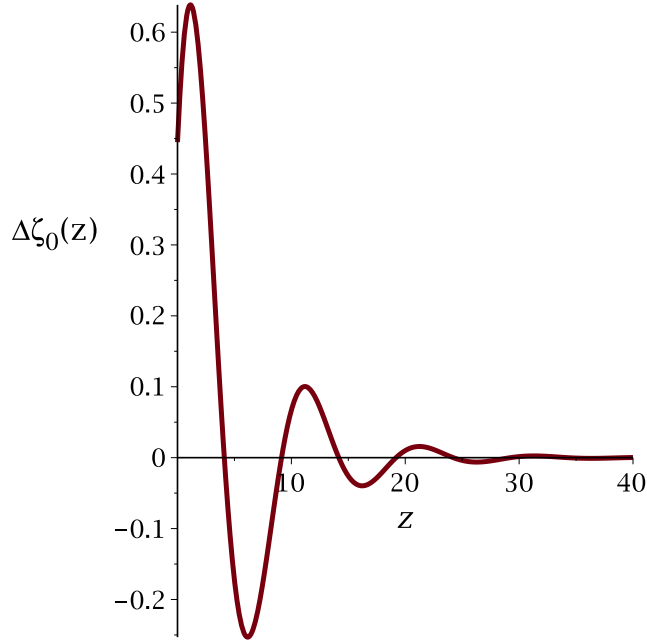}
\caption{The  excess volume-fraction (\ref{profile}) for the PY reference system (\ref{PY}) 
and the interaction potential (\ref{int_pot_r}) with
$K_1=1, K_2=0.2,\kappa_1=1,\kappa_2=0.5$. The thermodynamic state is given by $\tau=0.23$
with $\tau$ defined in Eq.(\ref{kappa}). 
$\Delta\zeta_0$ is in units of the dimensionless wall-particle attraction,  $\beta h$, and $z$ is in units of particle diameter.}
\label{fig_profile}
\end{figure}

  As discussed in sec.\ref{sec:cf},  the   effect of $\Delta\zeta_0$ 
  on the correlation function, neglected in (\ref{ELGmm}), should be smaller  for  $\zeta_b\approx 0.129$
   than for different volume fractions. Our choice of $\tau$ for $\zeta_b=0.129$ corresponds 
   to $T^*\approx 0.134$ (at the $\lambda$-line  $T^*\approx 0.126$ for $\zeta_b=0.129$).
 For the chosen interactions and $\tau$, the inverse decay length in the GM,
 $\alpha_0\approx 0.184$, is  very close to  $\alpha_0\approx 0.187$ obtained from Eq.(\ref{al}).
 The accuracy of $\alpha_1$ is not as good, $\alpha_1\approx 0.626$ and $\alpha_1\approx 0.576$ in 
 the GM and in Eq.(\ref{al}), respectively. When $\tau$ increases from $\tau=0$,  
 $\alpha_0$ in the GM and given by Eq.(\ref{al}) both increase, but $\alpha_1$  in the GM increases, whereas $\alpha_1$
 obtained from Eq.(\ref{al}) decreases. The GM becomes less accurate when the system becomes less inhomogeneous.  
 Still, given all the approximations, the GM works quite well compared to the
 linearized exact EL equation in the phase space region corresponding to inhomogeneities at a well-defined length scale. 
 The excess volume fraction in the GM is shown for   $\tau=0.23$ in Fig.\ref{fig_profile}.

\begin{figure}
  \includegraphics[scale =0.45]{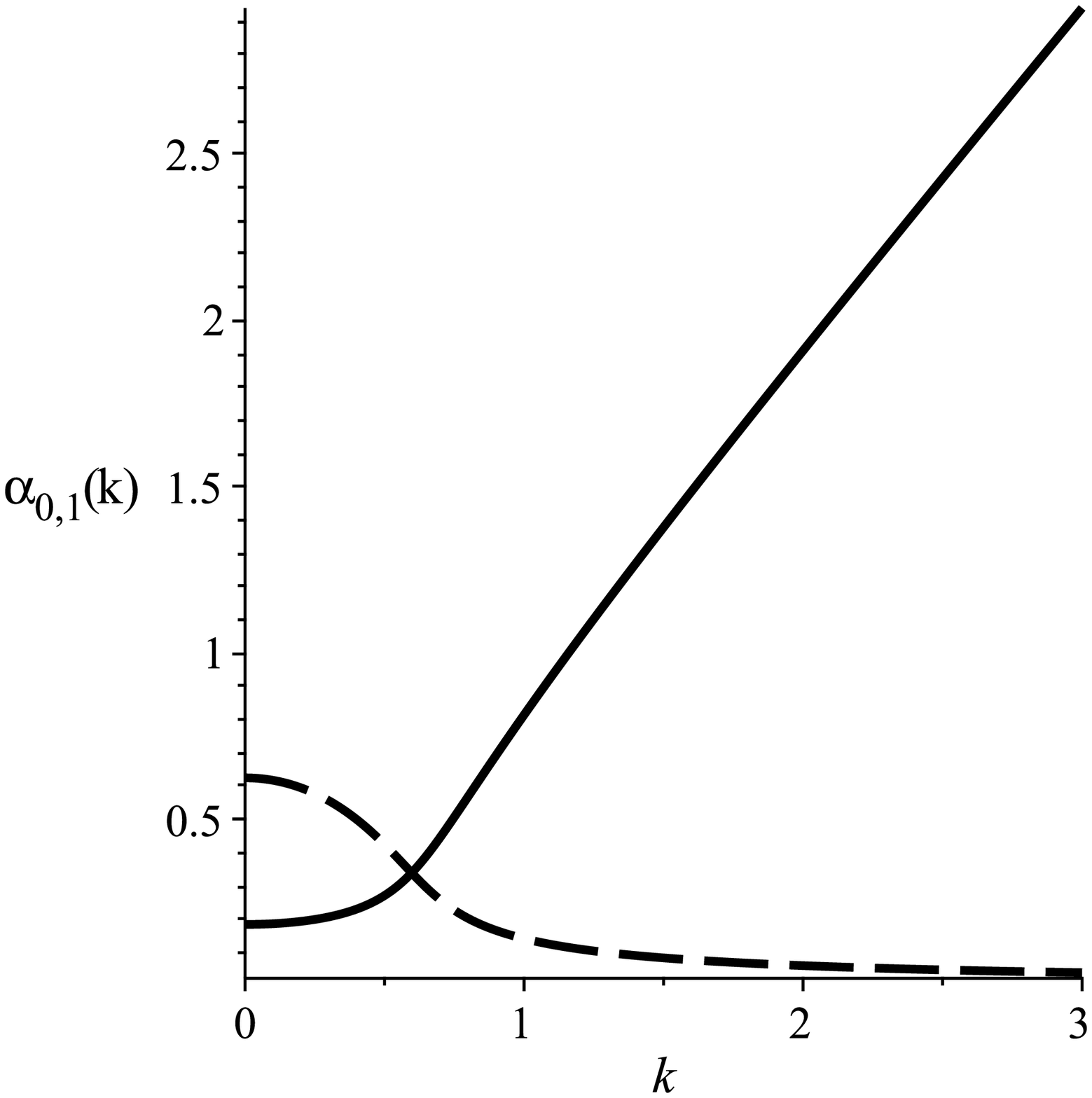}
\caption{ The inverse decay length $\alpha_0(k)$ (solid line) and  the wavenumber of the oscillatory
decay $\alpha_1(k)$ (dashed line)  in the direction perpendicular to the wall, of the correlations between 
the longitudinal $k$ modes (Eq.(\ref{alphakpar}) and (\ref{Gkzz})).
 The interaction potential is given in Eq.~(\ref{int_pot_r}) 
with $K_1=1, K_2=0.2,\kappa_1=1,\kappa_2=0.5$. The thermodynamic state is given by $\tau=0.23$
with $\tau$ defined in Eq.(\ref{kappa}). For this potential, $k_0\approx 0.59$ and the wavelength of the most probable
 density-wave in the bulk is $2\pi/k_0\approx 10$. 
 Volume-fraction waves in the longitudinal direction with $k\gg 2k_0$ correspond to disintegration of the aggregates.
 The decay length in the transverse direction of such fluctuations, $1/\alpha_0(k)$, is smaller than the particle diameter, i.e. 
they do not propagate to another layer of particles.}
\label{alfy}
 \end{figure}

In Fig.\ref{alfy} the inverse lengths characterizing the decay of the correlation between the longitudinal $k$-modes
in the planes at the distance $z=z_1$ and $z=z_2$ from the confining wall are shown. 
Note that the longitudinal density waves with 
the wavelengths larger than $2\pi/k_0$ excited in one plane decay much more slowly in the transverse direction 
than the short-wavelength longitudinal density oscillations. The short-wavelength longitudinal fluctuations in one plane
practically do not propagate to
different layers of particles. While long-wavelength density oscillations correspond merely to displacements, 
reorientation or reshaping 
of clusters or layers,
the density waves with the wavelengths shorter than $\pi/k_0$ correspond merely to disintegration of the aggregates. 

\begin{figure}
\includegraphics[scale =0.4]{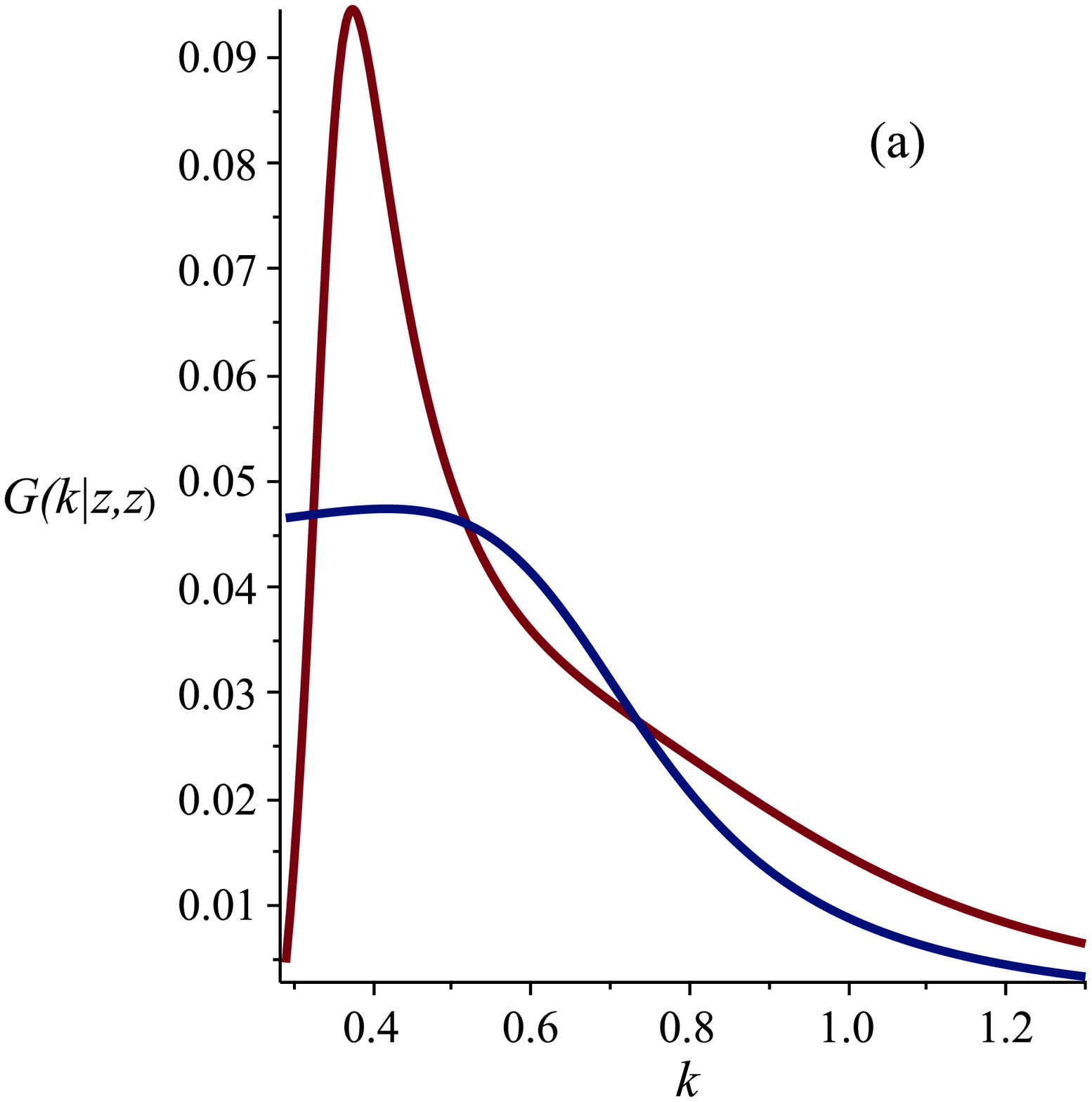}
\includegraphics[scale =0.4]{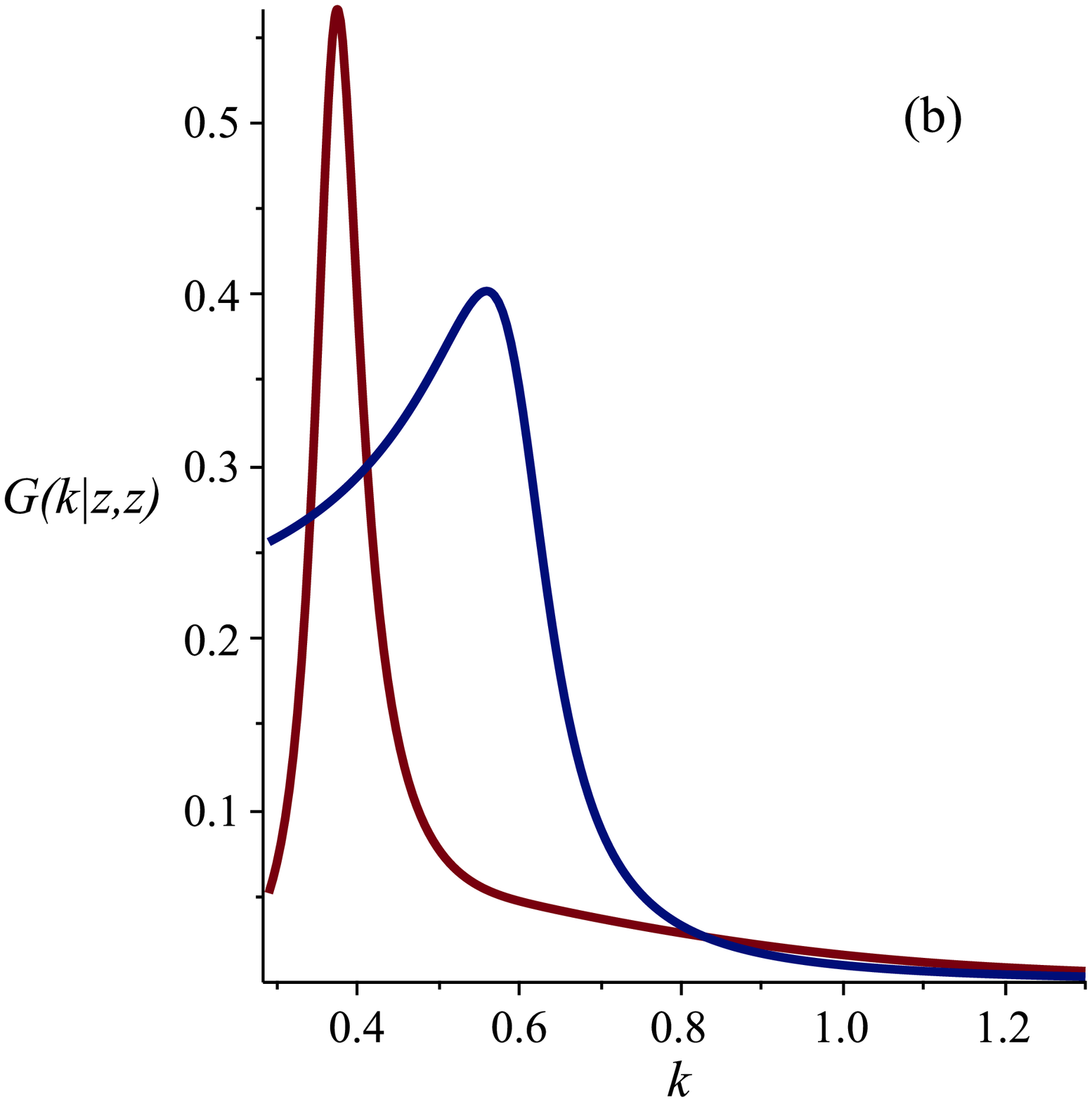}
\caption{The correlation function (\ref{Gkzz}) in 2D Fourier representation in the planes parallel to the wall
at $z=0$ (upper line),
 and $z=\infty$ (lower line) for the PY reference system 
and the interaction potential (\ref{int_pot_r}) with
$K_1=1, K_2=0.2,\kappa_1=1,\kappa_2=0.5$. The thermodynamic state is given by $\tau=0.316$ (a) and  $\tau=0.076$ (b)
with $\tau$ defined in Eq.(\ref{kappa}). The wavenumber is in units of inverse particle diameter.
}
\label{corfuk00}
\end{figure}

In order to describe the short-range order in the planes parallel to the wall, we present the structure factor for 
$z_1=z_2=z$, i.e  within planes parallel to  the wall. In Fig.\ref{corfuk00} we present $\tilde G(k_{\parallel}|z,z)$
for $z=0$ and $z=\infty$, far from and close to the $\lambda$-line. The shape of  $\tilde G(k_{\parallel}|z,z)$ 
follows from the fact that each term in (\ref{Gkzz}) has a maximum for
a different value of $k_{\parallel}$. The maximum of the first term is much broader than the 
maximum of the second term in (\ref{Gkzz}). Since the second term in (\ref{Gkzz}) vanishes for  $z\to\infty$,
this means significantly larger correlation length near the wall than in the bulk.
Both peaks are broader for larger $\tau$, indicating shorter correlation length
away from the $\lambda$-line, both in the bulk and near the wall, 
as expected. 
The position of the maximum for $z\to\infty$, $ k_{\parallel}=(k_0^2-\tau/\sqrt 3)^{1/2}$, depends on $\tau$ much more strongly 
than the position of the maximum for $z=0$. For large $\tau$, the maxima of 
$\tilde G(k_{\parallel}|0,0)$ and  $\tilde G(k_{\parallel}|\infty,\infty)$ occur for similar $k_{\parallel}$.
This means that  in the case of weak inhomogeneities, the wavelength of the volume-fraction oscillations near the wall
is similar to the wavelength of the oscillatory decay of correlations in the bulk.
The stronger are the inhomogeneities in the bulk, i.e. the smaller is the value of $\tau$, the larger 
is the difference between the period of oscillations near
the surface and in the bulk. 

\begin{figure}
\includegraphics[scale =0.45]{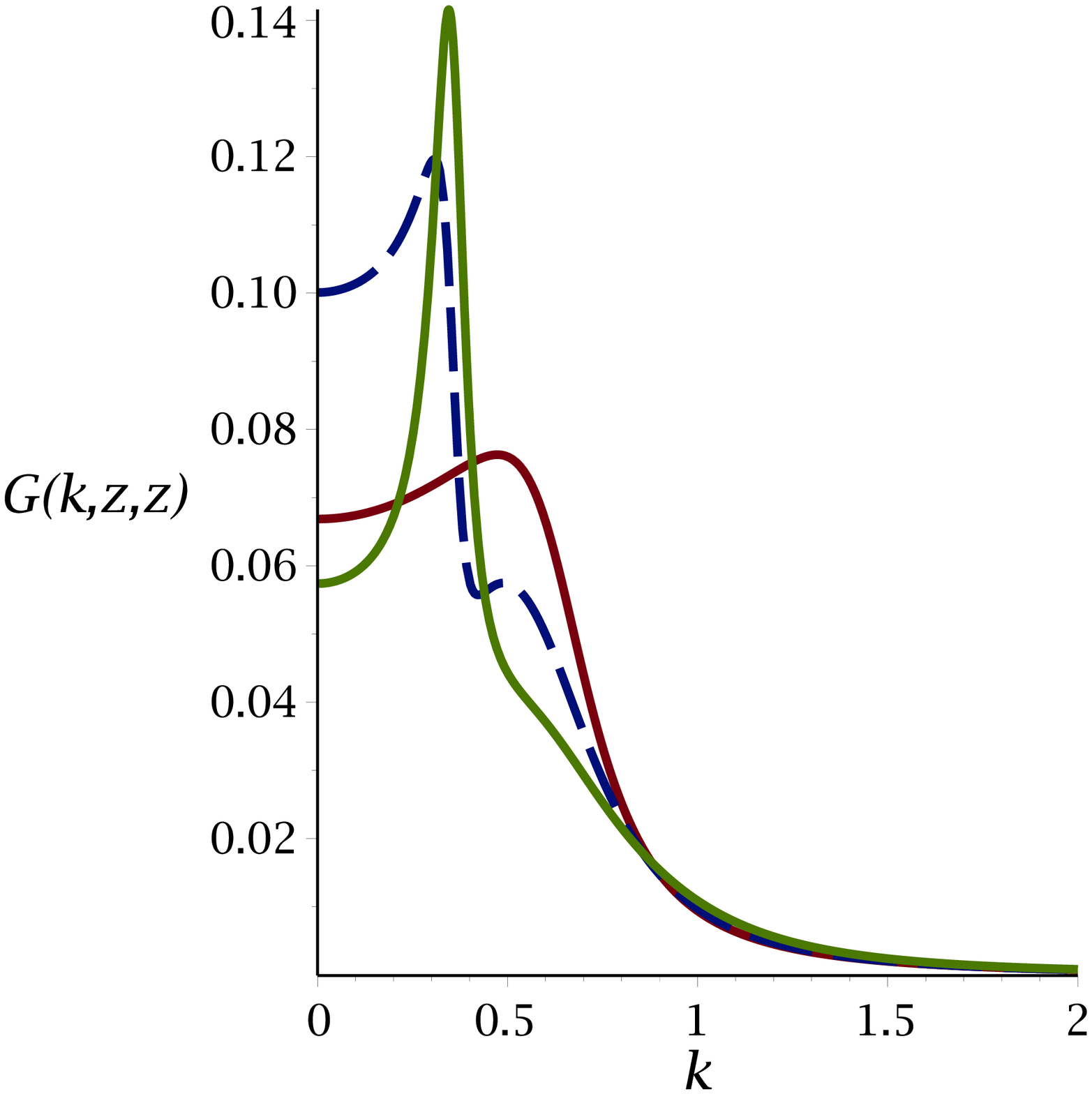}
\caption{The correlation function (\ref{Gkzz}) in 2D Fourier representation in the planes parallel to the wall
at $z=1$ (upper solid line),
$z=2$
(dashed line) and $z=\infty$ (lower solid line) for the PY reference system 
and the interaction potential (\ref{int_pot_r}) with
$K_1=1, K_2=0.2,\kappa_1=1,\kappa_2=0.5$. The thermodynamic state is given by $\tau=0.23$
with $\tau$ defined in Eq.(\ref{kappa}). The wavenumber is in units of inverse particle diameter.
}
\label{corfuk1}
\end{figure}

\begin{figure}
\includegraphics[scale =0.45]{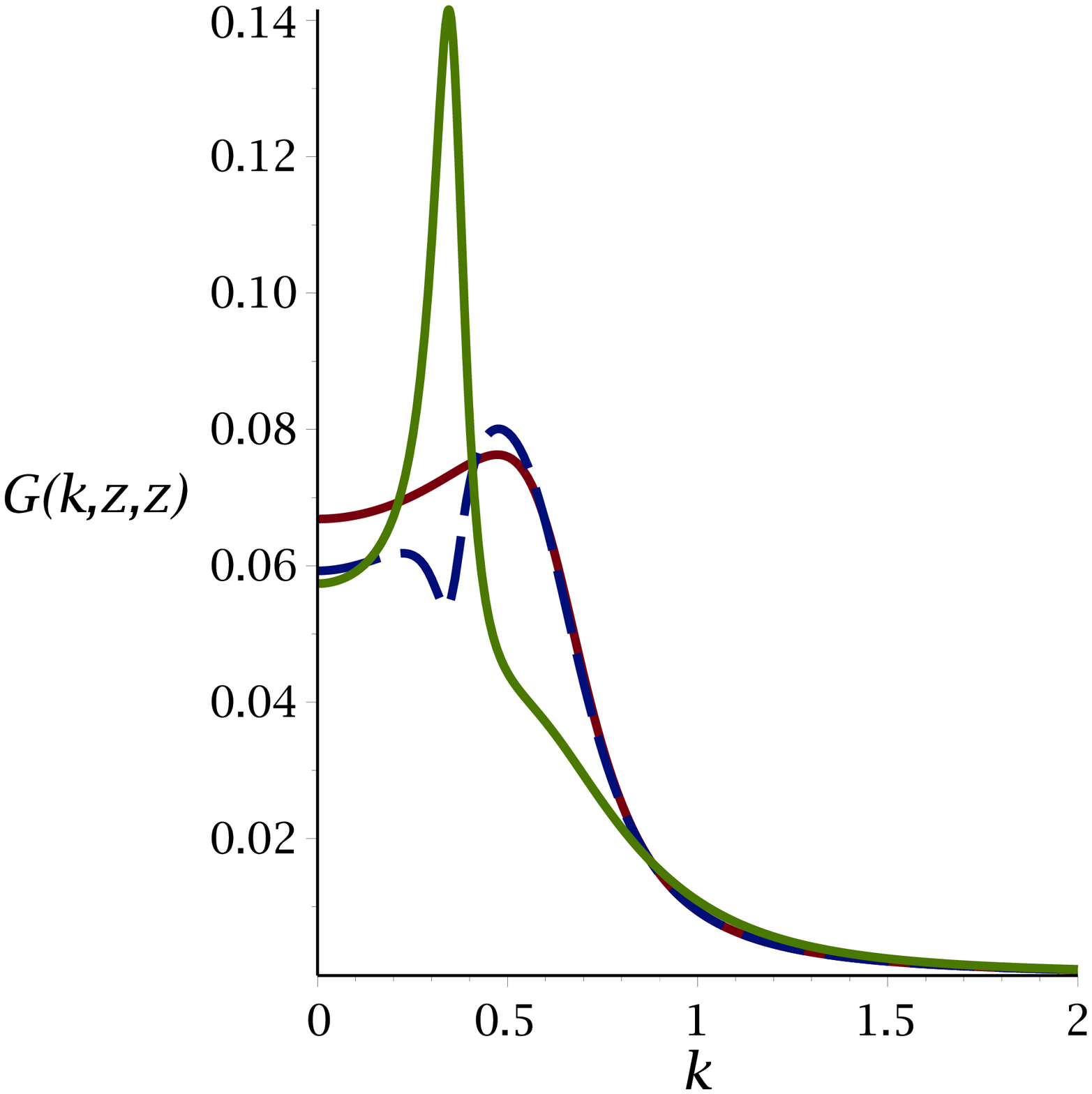}
\caption{The correlation function (\ref{Gkzz}) in 2D Fourier representation 
in the planes parallel to the wall at $z=1$ (upper solid line),
$z=4$
(dashed line) and $z=\infty$ (lower solid line) for the PY reference system and the interaction potential (\ref{int_pot_r}) with
$K_1=1, K_2=0.2,\kappa_1=1,\kappa_2=0.5$. The thermodynamic state is given by $\tau=0.23$
with $\tau$ defined in Eq.(\ref{kappa}).}
\label{corfuk2}
\end{figure}

\begin{figure}
\includegraphics[scale =0.45]{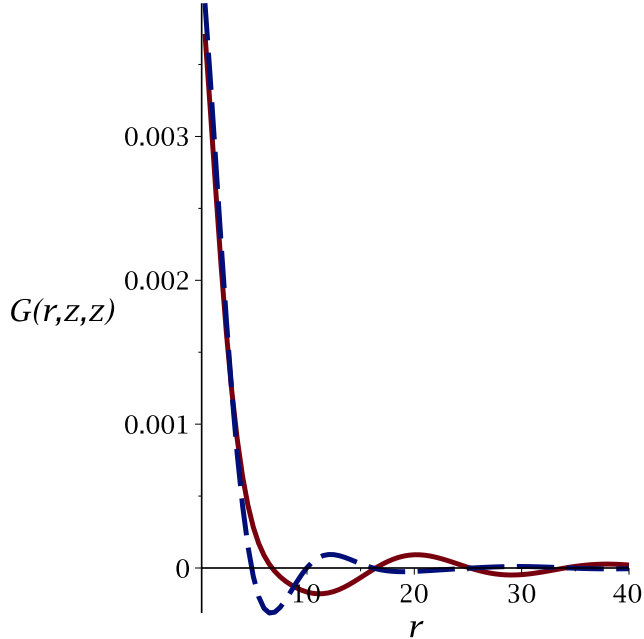}
\caption{The correlation function in real-space representation in the planes parallel to the wall at $z=1$ ( solid line),
and $z=4$
(dashed line)  for the PY reference system and the interaction potential (\ref{int_pot_r}) with
$K_1=1, K_2=0.2,\kappa_1=1,\kappa_2=0.5$. The thermodynamic state is given by $\tau=0.23$
with $\tau$ defined in Eq.(\ref{kappa}).}
\label{corfuk3}
\end{figure}

\begin{figure}
\includegraphics[scale =0.45]{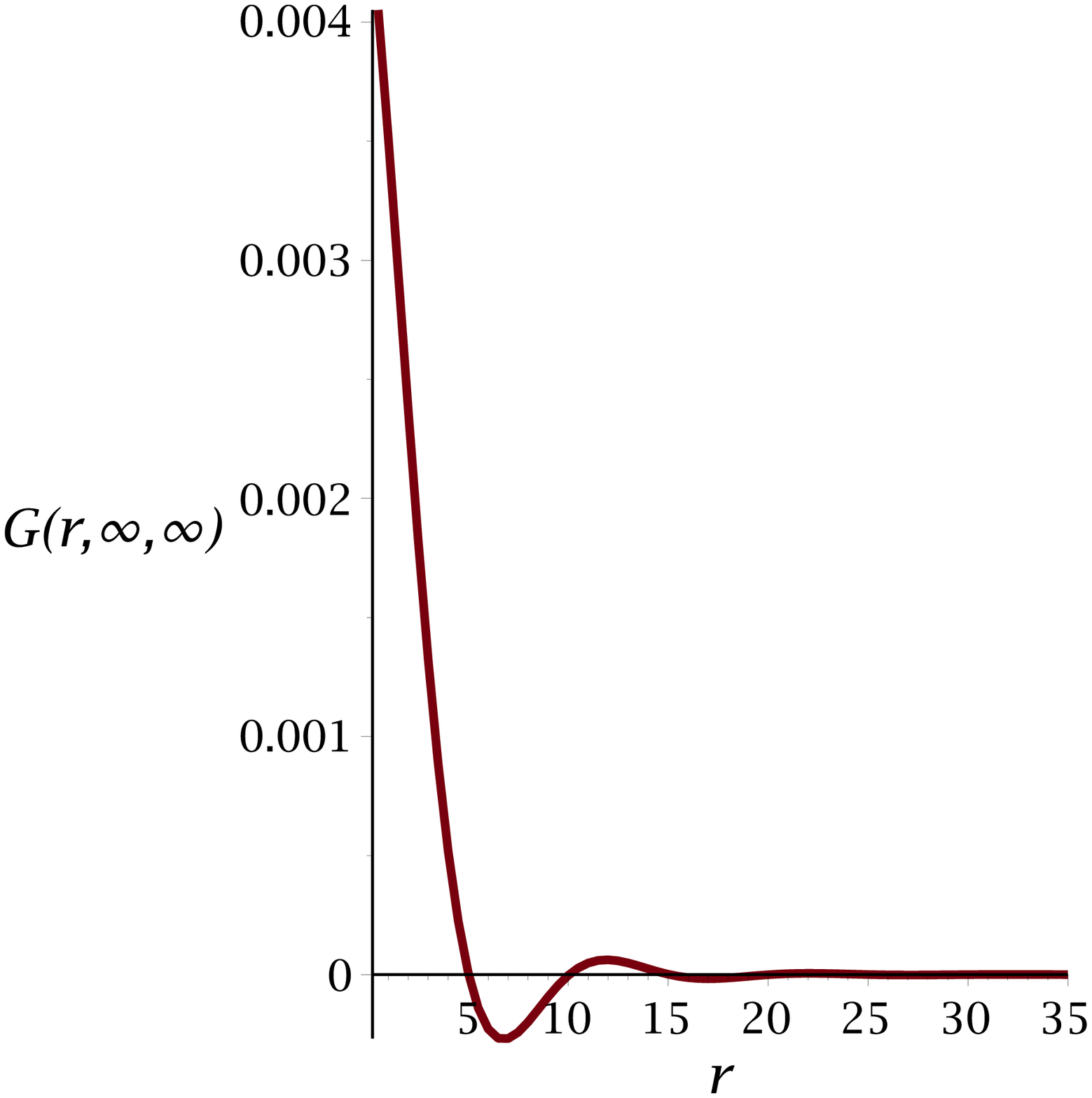}
\caption{The correlation function in real-space representation in the planes parallel to the wall at $z\to \infty$
for the PY reference system and the interaction potential (\ref{int_pot_r}) with
$K_1=1, K_2=0.2,\kappa_1=1,\kappa_2=0.5$. The thermodynamic state is given by $\tau=0.23$
with $\tau$ defined in Eq.(\ref{kappa}).}
\label{corfuk4}
\end{figure}

 In Figs.\ref{corfuk1} and \ref{corfuk2}  we present $\tilde G(k_{\parallel}|z,z)$ for $\tau=0.23$, and for 
 $z=1,2,4$ and $\infty$. 
  For this intermediate value of $\tau$, the maximum for $z\to\infty$ is much broader and occurs at significantly larger 
value of $k_{\parallel}$ than the maximum for $z\sim 1$. The range of the lateral periodic order is significantly
larger near the wall, and the period of oscillations is larger too (Figs.\ref{corfuk3},\ref{corfuk4}). 

As already discussed, we cannot expect accurate results for small $z$ in the Gaussian 
approximation that neglects the effect of the excess volume-fraction. 
The best accuracy is expected for volume fractions corresponding to formation of layers (lamellar phase) at low $T^*$.
Due to the missing-neighbors attraction to the wall, we may expect that the isotropic labyrinth of particle-rich region
in the bulk becomes anisotropic 
near the wall, with a tendency of the layers of particles to assume the parallel orientation.
Competition of this effect with the entropy leads to a larger wavelength of
the oscillatory decay of density correlations in the longitudinal direction near the wall than in the bulk. 
 
\section{\label{section:sum}Summary and discussion}

We have developed a mesoscopic theory for self-assembling systems near a confining surface.
We focused on the effects of the wall on a disordered inhomogeneous phase and limited ourselves to the MF approximation.
In the first step,  the standard DFT expression for the excess grand potential has been transformed 
to an equivalent form (Eq.(\ref{DbO})) that consists of the bulk and the surface contributions. The surface contribution
representing the missing neighbors beyond the confining surface (Eq.(\ref{bDUI})), 
is expressed in terms of moments of the interaction potential (Eq.(\ref{I2n+1})). 
Eq.(\ref{DbO}) allowed for a derivation of the 
EL equations for the volume-fraction profile and the correlation function in the near-surface region 
(Eq.(\ref{EL}) and (\ref{Dbm0})), with the BC following from Eq.(\ref{BC}).
%have been developed.  from the standard DFT expression for the excess grand potential.
%associated with the presence of the system boundary.
Solutions of these equations should be 
the same as the results of minimization of the postulated excess grand potential. 

In principle,  ordered periodic structure 
in the lateral direction could be induced in the vicinity of the confining surface, and we obtained equations for the 
excess volume fraction at the distance $z$ from the surface, $\Delta\zeta_0(z)$, and for the modulations of the volume fraction
in the planes parallel to the wall. In the rest of the paper we limited ourselves to the absence of the long-range order,
however. In simulations, only short-range periodic structure was found for the considered 
thermodynamic states~\cite{litniewski:19:0}.

We next considered various approximate versions of the theory, especially the linearized equation for $\Delta\zeta_0(z)$ 
(Eq.(\ref{Ulzl})) that  can be solved analytically. The analytical solution gives the asymptotic 
decay of the excess volume-fraction  at large distances. 

The lowest-order nontrivial approximation, GM, 
has been introduced in sec.\ref{gm}. It is based on the same approximation
for the interaction potential in Fourier representation  (Eq.(\ref{hatU}))
 as in the theory for bulk systems with mesoscopic inhomogeneities~\cite{ciach:13:0,ciach:18:0}.
 In addition, the series representing the missing-neighbors
 contribution to the excess grand potential associated with the 
 presence of the confining surface (Eq.(\ref{bDUI})) 
 is truncated at the first order term.  In this approximation, the missing neighbors contribution to the grand potential
 is proportional to  $\Delta\zeta_0(0)^2$. If in addition the wall-particle interaction is of very short range
 and we can assume a contact potential, the mathematical form of the GM resembles strongly Landau-type theory for a
 semiinfinite system, with the bulk part of the Brazovskii form. However, in our theory there are no
 free phenomenological parameters.
 All coefficients depend on the interaction potential and on the thermodynamic state. 
 
 Solutions of the linearized equations in GM are presented and discussed on a general level in 
 sec.\ref{sec:vfgm} and \ref{sec:cfgm}, and for a particular case of the double-Yukawa potential in sec.\ref{sec:example}. 
 The volume fraction profile has a form of exponentially damped oscillations, that very well reproduces results of simulations 
 except from $z<2\pi/\alpha_1$. The decay length and the wavenumber agree with simulation on a semiquantitative level.
 The GM quite well reproduces the solution of the more general equation (\ref{al}) for the decay length. 
 However, the wavenumber of oscillations deviates from the solution of (\ref{al}) 
 in the case of weak inhomogeneities (high $T^*$). The stronger the inhomogeneities, 
 the better the agreement between GM and Eq.(\ref{Ulzl}). 
 
 We have solved the equation for the correlation function only in GM and only in the Gaussian approximation.
 The correlation between volume fraction waves with the wavelength $k_{\parallel}$ in the planes at the distance $z_1$ and $z_2$ 
 from the wall (Eq.(\ref{Gkzz})) consists of two terms. The first one is a function of $|z_1-z_2|$ and is independent
 of the surface properties. This is a kind of  ``background'' bulk correlations, present for any distance from the wall.
 The second term is a function of $z_1+z_2$, 
 and depends on the missing-neighbors contribution. This term is significant only close to the wall. 
 Since the missing neighbors contribution depends on 
 $\int_0^{\infty} dz z \tilde V(k_{\parallel},z)$, the effect of the wall on the correlations depends
 on the shape of the interaction potential. 
 Both terms in Eq.(\ref{Gkzz}) exhibit oscillatory decay 
 with the same characteristic lengths that strongly depend on $k_{\parallel}$.
 The volume fraction fluctuations in longitudinal direction 
 with the wavelength shorter than the size of aggregates, $\sim \pi/k_0$, practically do not propagate to different layers.
 The largest range in the transverse direction (the same as the decay length of  $\Delta\zeta_0(z)$) have the  volume fraction
 fluctuations in the longitudinal direction with $k_{\parallel}=0$. 
 
 The short-range order in the planes parallel to the wall, described by $\tilde G(k_{\parallel}|z,z)$, has been investigated for
 the double-Yukawa potential, where we could obtain analytical results in the Gaussian approximation.
 Each term in $\tilde G(k_{\parallel}|z,z)$
 has a maximum for a different value of $k_{\parallel}$. The maximum of the bulk term is much broader,
 indicting shorter decay length. Larger decay length in the longitudinal direction near the surface than far from
 it was observed in simulations~\cite{litniewski:19:0,bildanau:19:0} in agreement with our predictions.

 We presented analytical  results for the simplest version of the theory.
 The analytical expressions allow to investigate asymptotic behavior and to draw general conclusions. 
 We hope that the solutions discussed above
 show the key properties of the near-surface structure of the disordered phase in self-assembling systems.
 Thanks to 
 the systematic derivation of various approximate versions of the theory, it is possible to obtain more accurate
 results for various model systems. The theory developed in this work can be a convenient tool for studies 
 of the ordering effects of external surfaces on systems with spontaneous inhomogeneities on the mesoscopic length scale.
 First of all, it will be interesting to solve the nonlinear equations in GM. 
 It is also of interest to extend the theory beyond  MF, by taking into account the fluctuation contribution 
 to the excess grand potential. To do so we shall generalize the theory 
 developed in Ref.\cite{ciach:18:0} to the semiinfinite
 system along the lines described in this work.

\section{Appendix. The parameters $k_0, \hat U(k_0^2), v$ and $I_1(k_{\parallel})$ for the
Double-Yukawa potential}
In Fourier representation the potential (\ref{int_pot_r}) takes the form
 \begin{eqnarray}
  \hat U(k^2)=4\pi \Big[
  \frac{K_2}{\kappa_2^2+k^2} -\frac{K_1}{\kappa_1^2+k^2}
  \Big].
 \end{eqnarray}
 The parameters in the Landau-Brazovskii type theory with $\hat U$ approximated by Eq.(\ref{hatU}) are 
 \begin{eqnarray}
  k_0^2=\frac{\kappa_1^2\sqrt{K_2} -\kappa_2^2\sqrt{K_1}}{ \sqrt{K_1}-\sqrt{K_2}},
 \end{eqnarray}
\begin{eqnarray}
  \hat U(k_0^2)=-4\pi \frac{(\sqrt{K_1}-\sqrt{K_2})^2}{\kappa_1^2-\kappa_2^2},
 \end{eqnarray}
 \begin{eqnarray}
  v=4\pi\frac{(\sqrt{K_1}-\sqrt{K_2})^4}{(\kappa_1^2-\kappa_2^2)^3\sqrt{K_1K_2}}.
 \end{eqnarray}

 The missing-neighbors interaction term (see (\ref{I2n+1})) can be easily calculated analytically, and the result is
 \begin{eqnarray}
  I_1(k_{\parallel})=2\pi\Big[
  \frac{K_2}{(\kappa_2^2+k_{\parallel}^2)^{3/2}}-\frac{K_1}{(\kappa_1^2+k_{\parallel}^2)^{3/2}}
  \Big].
 \end{eqnarray}

\section{Acknowledgements}
I would like to thank Vyacheslav Vikhernko for discussions, and Guillermo Zarragoicoechea, Ariel Meyra and Andres de Virgilis
for discussions and hospitality at the Instituto de Fisica de Liquidos y Sistemas Biologicos, UNLP in La Plata, 
where a part of this work was done.
 This project has received funding from the European Union Horizon 2020 research 
and innovation programme under the Marie
Sk\l{}odowska-Curie grant agreement No 734276 (CONIN).
An additional support in the years 2017-2020  has been granted  for the CONIN project by the Polish Ministry
of Science and Higher Education. 
Financial support from the National Science Center under grant No. 2015/19/B/ST3/03122 is also acknowledged.

% \bibliographystyle{prsty} 
%\bibliography{bibliography_18.bib}
\end{document}